\documentclass{dune}
\pdfoutput=1
\usepackage[pdftex,bookmarks,hidelinks]{hyperref}
\usepackage[utf8]{inputenc}
\usepackage{multicol}
\RequirePackage{cite}
\graphicspath{ {/graphics/} }
\let\OLDthebibliography\thebibliography
\renewcommand\thebibliography[1]{
  \OLDthebibliography{#1}
  \setlength{\parskip}{0pt}
  \setlength{\itemsep}{0pt plus 0.3ex}
}
\newif\ifdp
\newif\ifsp

\newcommand{\cut}[1]{}                          

\hypersetup{
    final=true,
    colorlinks=false,
    linktocpage=true,
    linkbordercolor=blue,
    citebordercolor=green,
    urlbordercolor=magenta,
    filecolor=black,
    pdfpagemode=UseOutlines,
    pdfborderstyle={/S/U},  
}

\begin{document}

\pagestyle{titlepage}

\date{}

\title{\scshape\LARGE \bf Electron Scattering and Neutrino Physics
\\
\vspace{10mm}
 \normalsize A NF06 Contributed White Paper \\
\vspace{5mm} 
Submitted to the Proceedings of the US Community\\ 
Study on the Future of Particle Physics (Snowmass 2021) \\
\vspace{10mm}
\vskip -10pt
}

\renewcommand\Authfont{\scshape\small}
\renewcommand\Affilfont{\itshape\footnotesize}

\author[1]{A.~M.~Ankowski$^{\thanks{Contributor}}\thanks{Currently at Institute of Theoretical Physics, University of Wroc{\l}aw, Wroc{\l}aw, Poland}$}
\author[2]{A.~Ashkenazi$^{*}$}
\author[3,4]{S.~Bacca$^{*}$}
\author[2,5]{J.~L.~Barrow$^{*}$}
\author[6]{M.~Betancourt$^{*}$}
\author[7]{A.~Bodek$^{*}$}
\author[8,9]{M.~E.~Christy$^{*}$}
\author[3]{L.~Doria$^{*}$}
\author[10]{S.~Dytman$^{*}$}
\author[1]{A.~Friedland$^{*}$}
\author[5]{O.~Hen$^{*}$}
\author[11]{C.~J.~Horowitz$^{*}$}
\author[12]{N.~Jachowicz$^{*}$}
\author[6]{W.~Ketchum$^{*}$}
\author[13]{T.~Lux$^{*}$}
\author[14]{K.~Mahn ${\thanks{Editor: mahn@msu.edu}}$}
\author[15]{C.~Mariani$^{*}$}
\author[16]{J.~Newby$^{*}$}
\author[6]{V.~Pandey ${\thanks{Editor: vpandey@fnal.gov}}$}
\author[17]{A.~Papadopoulou$^{*}$}
\author[18]{E.~Radicioni$^{*}$}
\author[19]{F.~S\'{a}nchez$^{*}$}
\author[3]{C.~Sfienti$^{*}$}
\author[20]{J.~M.~Ud\'{\i}as$^{*}$} 
\author[21]{L.~Weinstein$^{*}$\vspace{0.5 cm}}
\vspace{0.5 cm}

\author[22]{\\ L.~Alvarez-Ruso}
\author[23]{J.~E.~Amaro}
\author[24]{C.~A.~Arg{\"u}elles}
\author[25]{A.B. ~Balantekin}
\author[26]{S.~Bolognesi}
\author[6,27]{V.~Brdar}
\author[1]{P.~Butti}
\author[28]{S.~Carey}
\author[17]{Z.~Djurcic}
\author[29]{O.~Dvornikov}
\author[30]{S.~Edayath}
\author[6]{S.~Gardiner}
\author[6]{J.~Isaacson}
\author[5]{W.~Jay}
\author[31]{A. Klustov\'{a}}
\author[7]{K.~S.~McFarland}
\author[6]{A.~Nikolakopoulos}
\author[6]{A.~Norrick}
\author[32]{S.~Pastore}
\author[27]{G.~Paz}
\author[33]{M.~H.~Reno}
\author[22]{I.~Ruiz Simo}
\author[3]{J.~E.~Sobczyk}
\author[34]{A.~Sousa}
\author[1]{N.~Toro}
\author[35]{Y.-D.,~Tsai}
\author[6]{M.~Wagman}
\author[14]{J.~G.~Walsh}
\author[36]{G.~Yang\vspace{0.2 cm}}
\vspace{0.5 cm}

\affil[1]{SLAC National Accelerator Laboratory, Stanford University, Menlo Park, CA, USA}
\affil[2]{Tel Aviv University, Tel Aviv, Israel}
\affil[3]{Institut f\"{u}r Kernphysik and PRISMA$^{+}$ Cluster of Excellence, Johannes Gutenberg-Universit\"{a}t, 55128 Mainz, Germany}
\affil[4]{Helmholtz-Institut Mainz, Johannes Gutenberg-Universit\"{a}t Mainz, D-55099 Mainz, Germany}
\affil[5]{Massachusetts Institute of Technology, Cambridge, MA, USA}
\affil[6]{Fermi National Accelerator Laboratory, Batavia, IL, USA}
\affil[7]{University of Rochester, Rochester, NY, USA}
\affil[8]{Hampton University, Hampton, VA, USA}
\affil[9]{Thomas Jefferson National Accelerator Facility, Newport News, VA, USA}
\affil[10]{University of Pittsburgh, Pittsburgh, PA, USA}
\affil[11]{Center for Exploration of Energy and Matter and Department of Physics, Indiana University, Bloomington, IN, USA}
\affil[12]{Department of Physics and Astronomy, Ghent University, Proeftuinstraat 86, B-9000 Gent, Belgium}
\affil[13]{Institut de F\'isica d’Altes Energies (IFAE) - The Barcelona Institute of Science and Technology (BIST), Campus UAB, 08193 Bellaterra (Barcelona), Spain}
\affil[14]{Michigan State University, East Lansing, MI, USA}
\affil[15]{Center for Neutrino Physics, Virginia Tech, Blacksburg, VA, USA}
\affil[16]{Oak Ridge National Laboratory, Oak Ridge, TN, USA}
\affil[17]{Argonne National Laboratory, Lemont, IL, USA}
\affil[18]{INFN Sezione di Bari, via Amendola 173, Bari, Italy}
\affil[19]{University of Geneva, Section de Physique, DPNC, 1205 Gen\'{e}ve, Switzerland}
\affil[20]{Nuclear Physics Group and IPARCOS, Universidad Complutense de Madrid, 28040 Madrid, Spain}
\affil[21]{Old Dominion University, Norfolk, VA, USA}
\affil[22]{Instituto de F\'{i}sica Corpuscular, Consejo Superior de Investigaciones Cient\'{i}ficas  and Universidad de Valencia,  E-46980 Paterna, Valencia, Spain}
\affil[23]{Department of Atomic, Molecular and Nuclear Physics, University of Granada, E-18071 Granada, Spain}
\affil[24]{Department of Physics \& Laboratory for Particle Physics and Cosmology, Harvard University, Cambridge, MA, USA}
\affil[25]{University of Wisconsin, Department of Physics, Madison, WI, USA}
\affil[26]{IRFU, CEA, Universit\'e Paris-Saclay, Gif-sur-Yvette, France}
\affil[27]{Department of Physics \& Astronomy, Northwestern University, Evanston, IL, USA}
\affil[28]{Department of Physics and Astronomy, Wayne State University, Detroit, MI, USA}
\affil[29]{University of Hawaii, Honolulu, HI, USA}
\affil[30]{Department of Physics and Astronomy, Iowa State University, Ames, IA, USA}
\affil[31]{The Blackett Laboratory, Imperial College London, London SW7 2BW, UK}
\affil[32]{Department of Physics and the McDonnell Center for the Space Sciences at Washington University in St. Louis, MO, USA}
\affil[33]{Department of Physics \& Astronomy, University of Iowa, Iowa City, IA, USA}
\affil[34]{Department of Physics, University of Cincinnati, Cincinnati, OH, USA}
\affil[35]{University of California, Irvine, CA, USA}
\affil[36]{University of California, Berkeley, CA, USA}

\maketitle

\renewcommand{\familydefault}{\sfdefault}
\renewcommand{\thepage}{\roman{page}}
\setcounter{page}{0}
\pagestyle{plain} 
\clearpage

\textsf{\tableofcontents}

\renewcommand{\thepage}{\arabic{page}}
\setcounter{page}{1}
\pagestyle{fancy}

\fancyhead{}
\fancyhead[RO]{\textsf{\footnotesize \thepage}}
\fancyhead[LO]{\textsf{\footnotesize \rightmark}}

\fancyfoot{}
\fancyfoot[RO]{\textsf{\footnotesize Snowmass 2021}}
\fancyfoot[LO]{\textsf{\footnotesize Electron Scattering and Neutrino Physics}}
\fancypagestyle{plain}{}

\renewcommand{\headrule}{\vspace{-4mm}\color[gray]{0.5}{\rule{\headwidth}{0.5pt}}}


\section{Executive Summary}
\label{sec:exec_summary}





A thorough understanding of neutrino-nucleus scattering physics is crucial for the successful execution of the entire US neutrino physics program. Neutrino-nucleus interaction constitutes one of the biggest systematic uncertainties in neutrino experiments - both at intermediate energies affecting long-baseline Deep Underground Neutrino Experiment (DUNE), as well as at low energies affecting coherent scattering neutrino program - and could well be the difference between achieving or missing discovery level precision. To this end, electron-nucleus scattering experiments provide vital information 
to test, assess and validate different nuclear models and event generators intended to be used in neutrino experiments. 

As mentioned, for instance, in the DUNE Conceptual Design Report, ``uncertainties exceeding 1\% for signal and 5\% for backgrounds may result in substantial degradation of the sensitivity to CP violation and the mass hierarchy”~\cite{DUNE:2015lol}.  DUNE will use a highly capable set of ``near'' detectors to make precision neutrino-nucleus scattering measurements. Complementary probes\footnote{Pion scattering, for example, has been extensively used in current oscillation programs to reduce systematic uncertainties in conjunction with near detector data.}, such as precise electron-nucleus scattering data that can be obtained using controlled kinematics, provide data
to robustly test and 
resolve discrepancies in neutrino interaction physics.


Similarly, for the low-energy neutrino program revolving around the coherent elastic neutrino–nucleus scattering (CEvNS) physics at stopped pion sources, such as at ORNL, the main source of uncertainty in the evaluation of the CEvNS cross section is driven by the underlying nuclear structure, embedded in the weak form factor, of the target nucleus. To this end, parity–violating electron scattering (PVES) experiments, utilizing polarized electron beams, provide vital model-independent information of determining weak form factors. This information is vital in achieving a percent level precision needed to disentangle new physics signals from the standard model expected CEvNS rate.

Finally, it is important to emphasize that a necessary condition in addressing neutrino-nucleus interaction physics challenge is the close collaboration of nuclear physics (NP) and high-energy physics (HEP) communities. Electron scattering efforts that play vital role in the future of the neutrino program are inherently considered NP program while the neutrino programs are supported by HEP. Unfortunately, crossing the strictly defined boundaries between the NP and HEP funding agencies has been a significant challenge to progress.
It is also worth noting that the electron-nucleus and neutrino-nucleus scattering also provide complementary information on nuclear structure and dynamics that is interesting in its own right. A close collaboration between the two communities could help provide unique information on nuclear and hadronic physics. 

In this white paper, we highlight connections between electron- and neutrino-nucleus scattering physics at energies ranging from 10s of MeV to a few GeV, review the status of ongoing and planned electron scattering experiments, identify gaps, and layout a path forward that benefits the neutrino community. We also highlight the systemic challenges with respect to the divide between the nuclear and high-energy physics communities and funding that presents additional hurdle in mobilizing these connections to the benefit of neutrino programs.

\clearpage

\section{Introduction}
\label{sec:introduction}



The success of current and future neutrino experiments in achieving discovery level precision will greatly depend on the precision with which the fundamental underlying process -- neutrino interaction with the target nucleus in the detector -- is known~\cite{NuSTEC:2017hzk}. To this end, electron scattering experiments have been playing a crucial role by providing high-precision data as the testbed to assess and validate different nuclear models intended to be used in neutrino experiments. 

For the accelerator-based neutrino program, such as DUNE, the primary physics goals are determining mass hierarchy and  testing the three flavor paradigm with precision measurements, 
including subtle effects of $\delta_{CP}$~\cite{DUNE:2020jqi}. DUNE is a powerful facility with enormous physics reach, the intense neutrino source can also be used to probe for exotic physics, and the massive detectors can be used to measure neutrinos from astrophysical phenomena like supernova. Each of those physics thrusts relies on precision knowledge of neutrino-nucleus scattering.
The main challenges in constraining neutrino-nucleus scattering physics stems from the fact that neutrino energies at these experiments typically range from 100s of MeV to a few GeV where different interaction mechanisms yield comparable contributions to the cross section. One has to constrain an accurate picture of the initial state target nucleus, its response to the electroweak probe that include several reaction mechanisms resulting into various finals state particles, and final state interactions that further modify the properties of the hadronic system created at the primary interaction vertex. While electron and neutrino interactions are different at the primary vertex, many underlying relevant physical processes in the nucleus are the same in both cases, and electron scattering data collected with precisely controlled kinematics (initial and final energies and scattering angles), large statistics and high precision allows one to constrain nuclear properties and specific interaction processes. 

For the coherent elastic neutrino–nucleus scattering (CEvNS) program at stopped pion sources, such as at ORNL, the main source of uncertainty in the evaluation of the CEvNS cross section is driven by the underlying nuclear structure, embedded in the weak form factor, of the target nucleus. The recent detection of CEvNS process by the COHERENT collaboration~\cite{COHERENT:2017ipa} has opened up a slew of opportunities to probe several Beyond the Standard Model (BSM) scenarios in CEvNS experiments. In order to disentangle new physics signals from the SM expected CEvNS rate, the weak form-factor which primarily depends on the neutron density has to be known at percent level precision. Although, the ground state proton density distributions are relatively well constrained through elastic electron scattering datasets~\cite{DeVries:1987atn, Fricke:1995zz}, the neutron density distributions are only poorly known due to historic experimental difficulties in using electroweak probes such as polarized electron beams. Parity–violating electron scattering (PVES) experiments, utilizing polarized electron beams, provide relatively model-independent ways of determining weak form factors and neutron distributions of a target nucleus, that can be used as direct input to the CEvNS cross section.

For over five decades, the electron scattering experiments at different facilities around the world have provided wealth of information on the complexity of nuclear structure, dynamics and reaction mechanisms. Static form factors and charge distributions have been extracted from elastic scattering data, while measurements of inelastic cross sections have allowed for systematic studies of the dynamic response functions, which shed light on the role played by different reaction mechanisms. Decades of experimental work have provided vital testbed to assess and validate theoretical approximations and predictions that propelled the theoretical progress staged around~\cite{Benhar:2006wy, Benhar:2006er}. While previous and existing electron scattering experiments provide important information,  new
measurements to support neutrino-nucleus modeling are valuable. 
Dedicated electron scattering experiments with targets and kinematics of interests to neutrino experiments (CEvNS, supernova, and accelerator-based) will be vital in the development of neutrino-nucleus scattering physics modeling that underpin neutrino programs. 

This white paper is structured as follows. In Sec.~\ref{sec:impact-elec-neut}, we briefly outline the impact of electron scattering physics on neutrino programs. We then identify connections between electron- and neutrino-nucleus scattering physics in Sec.~\ref{sec:elec-to-neutrino}. In Sec.~\ref{sec:expt_landscape_acc-based}, we survey the current and planned electron scattering experiments that are input to accelerator-based neutrino program, and identify connections and gaps between them. In Sec.~\ref{sec:expt_landscape_low-energy}, we survey the current and planned electron scattering experiments that are input to CEvNS and supernova program. Thereby, in Sec.~\ref{sec:np-hep}, we highlight the challenges emerged from the divide between the nuclear and high-energy physics communities and funding that presents additional hurdle in mobilizing these connections and building a community. We summarize in Sec.~\ref{sec:conclusions}.


\section{Electron Scattering as a Vital Input to Neutrino Physics}
\label{sec:impact-elec-neut}

In this section, we briefly describe the neutrino interactions challenges faced by neutrino experiments in their key physics goals of measuring oscillations parameters, disentangling new physics signals and in detecting supernova neutrino signals. We then outline the key input that electron scattering experiments provide in tackling neutrino interactions modeling for both the long-baseline experiments as well as for CEvNS experiments. 

\subsection{Impact on Long-Baseline Oscillation Physics}









\subsubsection{Challenges of Interaction Modelling for LBL Programs}

We discuss the impact of electron scattering measurements to the future DUNE program, acknowledging that a similar set of arguments will apply for the Hyper-Kamiokande program. DUNE is a ``long baseline'' (LBL) neutrino experiment with 1) an intense, accelerator driven (anti)neutrino beam, 2) a suite of ``near'' detectors~\cite{DUNE:2021tad} close to the neutrino source production, and 3) a set of ``far'' detectors located in the neutrino beam. DUNE has a broad physics program, including precision measurements of neutrino oscillation physics parameters, searches for exotic physics, neutrino interactions and astrophysics.  The DUNE beam covers a wide range of energies (approximately 500 MeV to 5 GeV). Also, DUNE will collect interactions in the far detector from atmospheric (anti)neutrinos, which, when combined with beam data, provide enhanced sensitivity to oscillation physics. DUNE's beam intensity and detector size will provide unprecedentedly large samples of $\nu_e$/$\overline{\nu}_e$ appearance.  

To correctly estimate neutrino oscillation parameters of interest requires robust knowledge of neutrino interactions.  Neutrino oscillation parameters are inferred from event rates, which requires the correct mixture of processes. Across the DUNE beam energies, processes like quasielastic scattering, resonant production of pions, shallow inelastic scattering and deep inelastic scattering are all relevant. A prediction of each process, including the kinematics and multiplicity of outgoing particles is needed to estimate detection efficiency and therefore expected event rates. Oscillation occurs as a function of true neutrino energy, so the relationship between all final state particles and neutrino energy for each process is required. 

\begin{figure}
    \centering
    \includegraphics[width=0.99\linewidth]{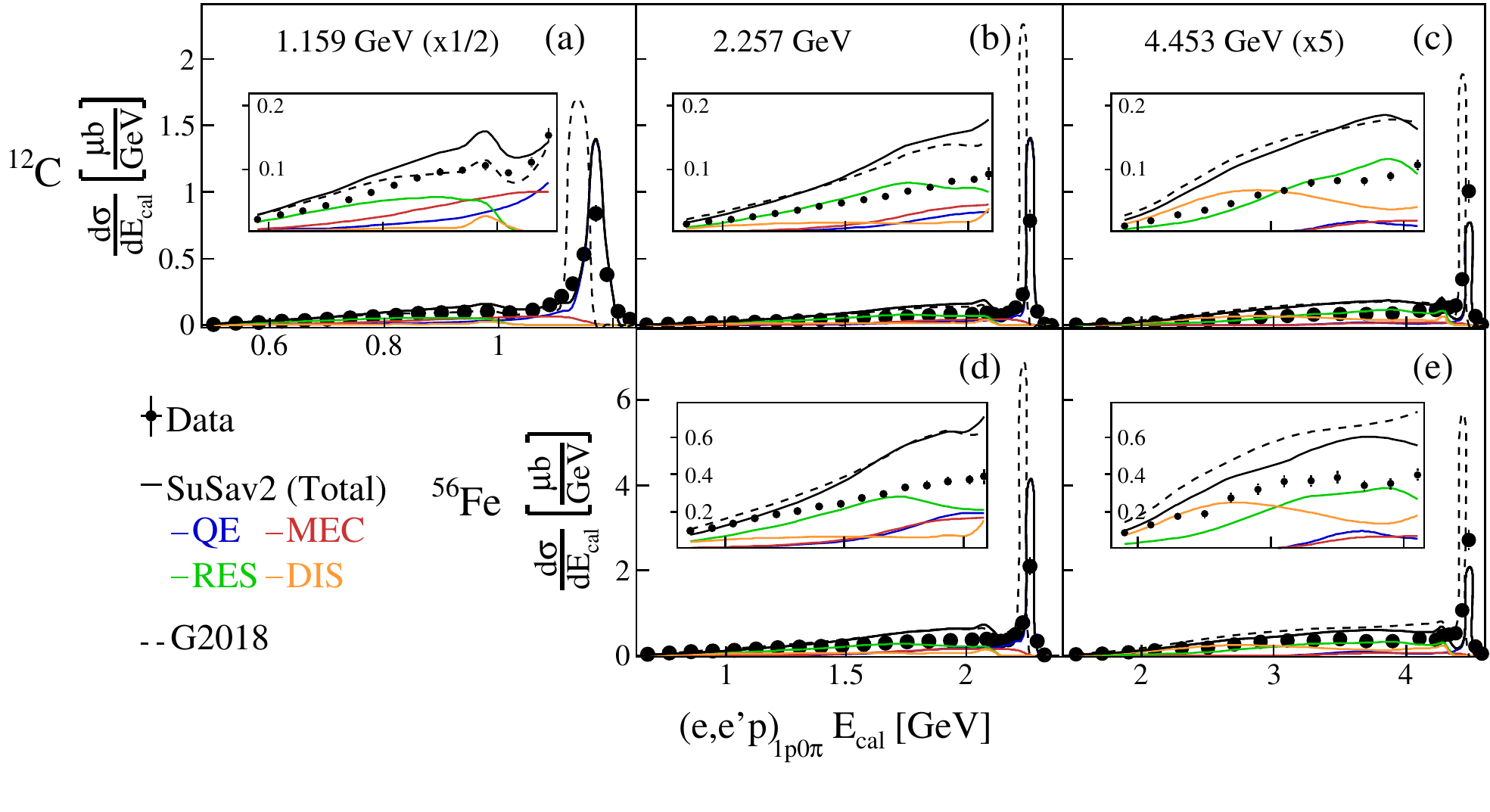}
    \caption{The $1p0\pi$ cross section plotted as a function of the reconstructed calorimetric energy Ecal for data (black points), SuSAv2 (black solid curve) and G2018 (black dotted curve) models. Different panels show results for different beam energy and target nucleus combinations: (top row) Carbon target at (left to right) 1.159, 2.257 and 4.453 GeV, and (bottom) Iron target at (left) 2.257 and (right) 4.453 GeV incident beam. The 1.159 GeV yields have been scaled by 1/2 and the 4.453 GeV yields have been scaled by 5 to have the same vertical scale. Coloured lines show the contributions of different processes to the GENIE SuSAv2: QE (blue), MEC (red), RES (green) and DIS (orange). Error bars show the 68\% confidence limits for the statistical and systematic uncertainties (not shown when they are smaller than data point). Figure from Ref.~\cite{{CLAS:2021neh}}.}
    \label{fig:e4nuerec}
\end{figure}

It is widely known that the predictions of interaction processes are not necessarily well modelled. Many models are a combination of a theoretically motivated model but often have empirical extrapolations in difficult to assess kinematic regions. Simulations of interaction models may also have to make approximations to make computations tractable. Interaction model relevant degrees of freedom are treated as nuisance parameters in oscillation fit. If this parameterization is incomplete or insufficient, this can also lead to issues in oscillation results. Notably, comparisons of experimental measurements of neutrino processes at different energies do not present a consistent or clear picture as to the origin of inconsistencies. As is summarized in Ref~\cite{NuSTEC:2017hzk}, experimental and theoretical groups have reported the possibility of bias in oscillation results due to deficiencies in the interaction model.  One projection of this is shown in Fig.~\ref{fig:e4nuerec}, from Ref.~\cite{CLAS:2021neh}, which shows the cross section as a function of the reconstructed energy (Ecal) for 1.159, 2.257 and 4.453 GeV carbon events and 2.257 and 4.453 GeV iron events.  The reconstructed neutrino energy spectra show a sharp peak at the real beam energy, followed by a large tail at lower energies. For carbon, only 30–40\% of the events reconstruct to within 5\% of the real beam energy. For iron this fraction is only 20–25\%, highlighting the crucial need to well model the low-energy tail of these distributions. This disagreement in the tails increases with energy and nuclei number. 

Oscillation experiments mitigate these issues partially with data from a ``near detector'', positioned close to the neutrino beam origin to constrain interaction physics. DUNE's design includes a movable set of precision detectors with broad acceptance of particle composition and kinematics. The detector motion (``PRISM'') feature will enable, within the same detector, comparisons of event rates from neutrino energy spectra which are different. This will provide information, akin to electron scattering, in order to mitigate interaction model bias.

Electron scattering data is complementary to the information on the cross sections from the near detectors. First, although the near detector measures both the axial and the vector part of the cross-section, the vector part, precisely provided by electron scattering must be correctly predicted for the right inference in predicting asymmetries in the neutrino-antineutrino rate, key for measurements like CPV. Second, the near detector does not have uniform or complete acceptance. Therefore, predictions at the DUNE far detector still rely on a prediction which depends on particle multiplicity and kinematics. Third, detectors have finite thresholds\footnote{The use of the ND-GAr detector aims to provide a lower threshold than is used at the far detector to test models.}, and some particles, like neutrons, may be challenging to associate to a given neutrino interaction. Fourth, robust interaction models are necessary for a variety of exotic physics searches which use the near detector to search for signal, and therefore do not benefit from the constraint on the cross section the near detector provides in oscillation analyses. Finally, information from electron scattering is used to develop theoretical models in advance of early data from DUNE near detectors.

\subsubsection{Unique Benefits of Electron Scattering for LBL}

Given the inherent complexity of the neutrino-nucleus scattering problem, thorough validation of models against data is required. It has long been recognized that a robust model must reproduce not only neutrino-scattering data, but also measurements of electron-nucleus scattering ~\cite{Smith:1972xh}. While there are several important differences between neutrino and electron scattering, the two processes share many common elements. These range from modeling of the nuclear ground state, to transport of final-state hadrons through the nucleus, to the primary-vertex interaction physics in the deep-inelastic limit (see, e.g., Ref.~\cite{Ankowski:2019mfd} for discussion). By exploiting electron-scattering data to fix the common physics elements, one will then be able to isolate the ingredients that are specific to neutrino interactions, such as axial couplings.

The idea of using electron data was further developed and strengthened since the turn of the millennium~\cite{Miller:2002sj,Gallagher:2004nq}. 
It was incorporated into the mission statement of the GENIE generator, was mentioned prominently in the original paper~\cite{Andreopoulos:2009rq}, and comparisons at low energies were presented in Ref.~\cite{Katori:2013eoa}.
Likewise, the generator GiBUU was tested against electron-scattering data~\cite{Buss:2007ar,Leitner:2008ue,Gallmeister:2016dnq,Mosel:2017anp,Mosel:2018qmv}.

What makes comparisons with electron-scattering data such a powerful tool is the precise control of the scattering kinematics it affords. Neutrino-scattering measurements are by necessity reported integrated over a range of energies and angles. The integrated kinematics may mask significant problems, which may reappear on different targets or in different kinematic regimes in future analyses. In contrast, by precisely controlling scattering parameters, electron-scattering experiments are able to zero in on specific physical processes. 
 
\begin{figure}
    \begin{center}
        \includegraphics[width=0.99\textwidth]{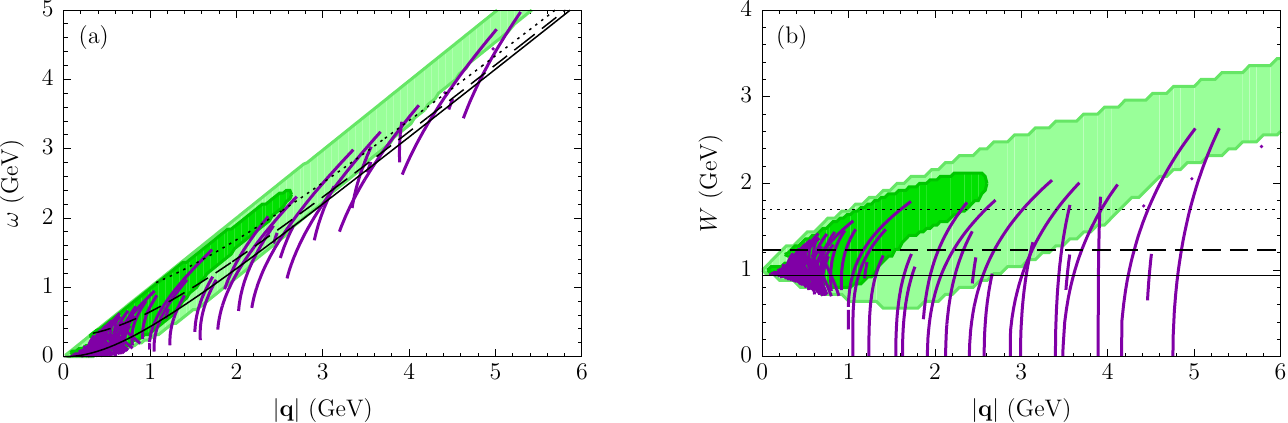}
       \caption{Kinematic coverage of available data for inclusive electron scattering on carbon~\cite{Whitney:1974hr,Barreau:1983ht,O'Connell:1987ag,Bagdasaryan:1988hp,Baran:1988tw,Sealock:1989nx,Day:1993md,Arrington:1995hs,Arrington:1998hz,Arrington:1998ps,Fomin:2008iq,Fomin:2010ei,Dai:2018xhi}, the most extensively studied nucleus, overlaid on the charged-current $\nu_\mu$Ar event distribution expected for the 2020 flux estimate~\cite{DUNE:2021cuw} in the DUNE near detector according to GENIE 3.0.6. The thick solid lines representing the measured cross sections are shown in the (a) $(|\mathbf{q}|, \omega)$ and (b) $(|\mathbf{q}|, W)$ planes, with momentum transfer $\mathbf{q}$, energy transfer $\omega$, and hadronic mass $W$.
       The dark and light shaded areas correspond to 68\% and 95\% of the expected events. The thin solid, dashed, and dotted lines present the kinematics corresponding to quasielastic scattering, $\Delta$ excitation, and the deep-inelastic scattering at $W = 1.7$ GeV on free nucleons. Only scarce information on the cross sections is available in the whole deep-inelastic region (above the thin dotted lines) and for higher-resonance production at $|\mathbf{q}| \gtrsim 1$ GeV (above the thin dashed lines) in the region of most importance for DUNE. For argon, a single dataset is currently available~\cite{Dai:2018gch}.}
        \label{fig:data_coverage}
    \end{center}
\end{figure}

The utility of validating a~neutrino Monte Carlo generator against electron-scattering data that it had not been tuned to was demonstrated recently in Refs.~\cite{Ankowski:2020qbe,electronsforneutrinos:2020tbf}. Figure~\ref{fig:data_coverage} shows the kinematic coverage of available data for inclusive electron scattering on carbon that was used in Ref.~\cite{Ankowski:2020qbe}. This coverage is superimposed on the expected distribution of the charged-current $\nu_\mu$Ar events in the DUNE near detector. The availability of this data made it possible to undertake a systematic study benchmarking GENIE v2.12 predictions across the different scattering regimes relevant to NOvA and DUNE. Identifying which regimes host largest discrepancies is the first necessary step towards the program of improving generator physics.

One notable finding from the comparisons at multi-GeV energies is a pattern of large discrepancies in the inelastic regime, including the region of higher resonances and its transition to deep inelastic scattering (the RES-to-DIS region). This pattern reflects an inherent difficulty of the physics problem, where no first-principles treatment exists and prescriptions based on various physics models must be combined in ways that are consistent with quark-hadron duality and reproduce the available data. The large discrepancies exist not only in inclusive cross sections, but also in exclusive variables~\cite{Ankowski:2019mfd}.

The physics uncertainties in the RES-to-DIS regime may have a profound impact on the performance of DUNE. Indeed, more than half of neutrino interactions there are expected to occur in this regime and the resulting rich hadronic final states will need to be reliably modeled to reconstruct the energy of the incoming neutrino. This especially concerns the energy fraction carried out by multiple neutrons created at various stages of the event development, as these are very difficult to reconstruct~\cite{Friedland:2018vry,Friedland:2020cdp}.

It is notable in this regard that recent measurements by MINERvA report discrepancies between the Monte Carlo predictions and pion-production data, and that no tune could be found that would reconcile all of the channels in the data~\cite{MINERvA:2019kfr}. This puzzle may be difficult to disentangle with neutrino data only, but corresponding electron scattering studies could help isolate the source of the problem.

To develop and validate improved models of neutrino-nucleus scattering in the RES-to-DIS will require new, high-precision electron-scattering data. This is already evident from the right panel of Fig.~\ref{fig:data_coverage}, which reveals insufficient coverage for large values of the hadronic mass $W$. There is therefore an urgent need for new measurements that include both the outgoing electron and the final-state hadronic system. In Sec.~\ref{sec:expt_landscape_acc-based}, we discuss a new program of this type at Jefferson Laboratory ($e4\nu$), and outline the capabilities of LDMX, an experiment proposed at SLAC.

Electron scattering measurements are also used as input to improved theoretical modelling. In discussions at a Snowmass workshop, there was strong demand for semi inclusive measurements of $(e,e^{\prime} p)$; $(e,e^{\prime} \pi)$ scattering for neutrino modelling.  There is also a lack of data on deuterium or neutron targets.  Finally, for both theorist and experimentalists, new projections of kinematic variables of interest is extremely important for model building and validation, as are updated and detailed systematic uncertainty assessments.


\subsection{Applications to Searches for Exotica and Nuclear Physics}

DUNE also has significant physics reach in exotic physics. DUNE will search for proton decay, light dark matter, heavy neutral leptons, and sterile neutrinos. Each of these searches, like the oscillation program, relies on (anti)neutrino rates. In some cases, near detector data will be able to constrain the interaction model, but not always, especially for searches performed using the near detector where any measurement of the signal process may be complicated by new physics. Searches for signals similar to neutral current interactions directly benefit from electron scattering measurements, and which are sometimes notorious to measure in neutrino detectors.



DUNE will also be able to make new measurements of interest to nuclear physics, which are enhanced by electron scattering. The DUNE near detector complex, by virtue of sampling multiple energy spectra, can effectively make pseudo-monochromatic fluxes akin to electron scattering~\cite{DUNE:2021tad}. Then, by comparing neutrino and antineutrino rates, DUNE can then produce measurements of  the axial-vector interference terms. There are no studies of the axial response at DUNE's energies. These unique measurements will be improved by correctly accounting for the vector piece of the cross section provided by electron scattering.




\subsection{Impact on Low-Energy Neutrino Physics}


\subsubsection{Coherent Elastic Scattering} 
\label{sevens}

In 1974 Fredman suggested that "if there is a weak neutral current, then the elastic process $\nu+A \rightarrow \nu + A$ should have a sharp coherent forward peak just as the  $e+A \rightarrow e + A$ does"~\cite{Freedman:1973yd}. Due to the elusive nature of neutrinos and the necessary technological advancement needed to detect such a tiny signal, 
it took 43 years for the first  coherent elastic neutrino scattering (CEvNS)  to be experimentally observed by the
COHERENT collaboration~\cite{COHERENT:2017ipa}. A stopped-pion
neutrino beam from the Spallation Neutron Source at
Oak Ridge National Laboratory impinged on a sodium-doped CsI
detector, allowing for an experimental discovery of CEvNS at the
$6.7\sigma$ confidence level.
A few years later, the COHERENT collaboration used a 24-kg active-mass liquid-argon scintillator detector~\cite{COHERENT:2020iec} and found that two independent analyses preferred CEvNS over the background-only null hypothesis with a significance larger than $3\sigma$.

Coherent elastic neutrino scattering off
a nucleus occurs in the regime $qR \ll 1$, where $q=|{\bf q}|$ is the
absolute value of the three-momentum transfer from the neutrino to the
nucleus, and $R$ is the nuclear radius.  
The  CEvNS cross section is
\begin{eqnarray}
\label{eq_cs}
\frac{d\sigma}{dT}(E_\nu,T) &\simeq& \frac{G_F^2}{4\pi} M \left[ 1- \frac{MT}{2E_\nu^2} \right] Q^2_W F^2_W(q^2) \,,
\end{eqnarray}
where $G_F$ is the Fermi constant, $M$ the mass of the nucleus, $E_\nu$
and $T$ the energy of the neutrino and  the nuclear recoil energy, respectively. The weak form factor $F_W(q^2)$ and weak charge  $Q_W$ are given by
\begin{eqnarray}
\label{eq_w}
 F_W(q^2) &=& \frac{1}{Q_W}[ NF_n(q^2) -(1-4\sin^2\theta_W)ZF_p(q^2) ] \\ 
 \nonumber
Q_W &=& N -(1-4\sin^2\theta_W)Z \,, 
\end{eqnarray}
where $\theta_W$ is the Weinberg mixing angle.
In Eq.~(\ref{eq_w}) the quantities
 $F_{p}(q^2)$ and $F_{n}(q^2)$ are the proton ($p$) and neutron ($n$) form factors,  respectively.
 While the proton distributions are relatively well known through elastic electron scattering experiments~\cite{DeVries:1987atn, Fricke:1995zz, Ottermann:1982kr}, neutron distributions are much more difficult to constrain.
 Because  $1-4\sin^2\theta_W(0) \approx 0$, on the one hand the weak
 form factor becomes $F_W(q^2)\approx F_n(q^2)$, and on the other hand the cross section scales like ${d\sigma}/{dT} \approx N^2$ (in the limit of $q \rightarrow 0$).
 So far, data on CsI and Ar established the $N^2$ scaling of the cross section and, using some phenomenological parameterization of the nuclear form factors, obtained a  
 signal that is consistent with the standard model~\cite{COHERENT:2020iec}. 
 Measuring CEvNS with increased precision opens up the possibility to discover new physics beyond the standard model, such as non  standard neutrino interactions~\cite{Davidson:2003ha,Barranco:2005yy, Liao:2017uzy}. For this reason, several experiments are being performed, or planned for,  to measure CEvNS with increased precision using a variety of detector technologies and neutrino beams that come either from stopped pions or from reactors,  see Table~\ref{tab:CEvNS}.
 
\begin{table}[tb]
\centering
\caption{Worldwide efforts in measuring CEvNS. List of planned experiments. \\}
\label{tab:CEvNS}
\begin{tabular}{ c | c | c | c }
 \hline
  Experiment & Technology & Location & Source \\
  \hline
  COHERENT &CSI, Ar, Ge, NaI & USA&$\pi$DAR\\
  CCM &Ar &USA&$\pi$DAR \\
  CONNIE &Si CCDs&Brazil & Reactor \\
  CONUS& HPGe&Germany& Reactor\\
  MINER&Ge/Si cryogenic& USA& Reactor\\
  NuCleus& Cryogenic CaWO$_4$, Al$_2$O$_3$ calorimeter array& Europe& Reactor\\
  $\nu$GEN &Ge PPC&Russia & Reactor\\
  RED-100 & LXe dual phase& Russia& Reactor \\
  Riochet & Ge,Zn & France & Reactor\\
  TEXONO &p-PCGe& Taiwan& Reactor \\
  NCC-1701 &p-PCGe& Germany& Reactor \\
\hline    
\end{tabular}
\end{table} 

In order to disentangle new physics signals from the SM expected CEvNS rate, the weak form factor, which primarily depends on the neutron density, has to be known at percent level precision.
Nuclear theory calculations can be used  to obtain the weak form factors.
Various theories based on the use of proton 
and neutrons as degrees of freedom have recently computed weak form factors for selected nuclei from $^{12}$C  to $^{208}$Pb~\cite{Payne:2019wvy, Yang:2019pbx, Hoferichter:2020osn, Co:2020gwl, VanDessel:2020epd}, using  different schemes that range from the mean field approximation to the shell model and the ab initio methods.
In particular for $^{40}$Ar, several many-body methods have been used
and even different nuclear Hamiltonians have been implemented within the same many-body method (see e.g., Ref.~\cite{Payne:2019wvy}), leading to an estimate of the size of nuclear structure effects of just a few percent in the low-$q$ regime relevant for CEvNS. Knowing that
nuclear structure uncertainties are small is a very promising information for beyond standard model searches with CEvNS.
In the future, more efforts will be devoted towards theoretical calculations in order to assess nuclear structure corrections in nuclei other than
$^{40}$Ar. Every theoretical calculation should be tested against elastic electron scattering  to prove that the proton distributions are well described.  Finally, it will be exciting to incorporate beyond standard model physics into modern nuclear structure calculations for CEvNS.

\subsubsection{10s of MeV Inelastic Scattering}

Stars with a mass larger than 8 times that of the Sun undergo  a core collapse process, where  99$\%$ of the 
gravitational binding energy of the inner proto-neutron star is shed within just 10 second of time into neutrinos  and antineutrinos  of all flavors. 
The kinetic energy of the produced neutrino/antineutrinos is maximally  50 MeV.
Because core-collapse supernova  are still not well understood, the detection of supernova neutrino from an event in our Galaxy could dramatically influence our understanding  of such astrophysical objects. Furthermore, even though they are rare events, in the whole Universe thousands of core-collapse explosions occur every hour,  resulting in a diffuse supernova neutrino background. The latter has not yet been seen, but an observation would greatly enhance our understanding of core-collapse supernovae.

Solar neutrinos have already enabled very important discoveries, from the  neutrino oscillation~\cite{SNO:2001kpb} that solved the solar neutrino problem~\cite{Cleveland:1998nv} to the first detection of CNO neutrinos from the Sun~\cite{BOREXINO:2020aww}. Still, many of their facets need to be unveiled. For example, we would like to know whether solar and reactor neutrinos oscillate with the same parameters. Solar neutrinos, mostly generated though the hydrogen burning reactions occurring in the center of our Sun, are of electron kind and reach  up to  15 MeV of energy in case of the $^8$B neutrinos.

Experimentally, one can use the charge current inelastic  process to detect  supernovae and solar neutrinos. For the supernovae neutrinos, multiple detectors are  set up  worldwide and are all connected  via an early warning system~\cite{SNEW}.  For the solar neutrinos, DUNE~\cite{DUNE:2015lol} plans to measure them with increased precision using the   $\nu_e + ^{40}{\rm Ar} \rightarrow e^- + ^{40}{\rm K}^*$ process.

Inelastic interactions of neutrinos with nuclei are still poorly understood:  theory is sparse and experiments have large error bars.
Because inelastic neutrino interactions have big uncertainties, in the future it will be crucial to measure inelastic electron scattering cross sections at energies below the 50 MeV mark and use those data to calibrate theoretical models for the neutrino scattering process. From the theory side, inelastic scattering has been recently tackled with mean-field methods~\cite{VanDessel:2020epd} and further efforts will need to be devoted to compute these  quantities from other theories, including ab initio nuclear theory. Overall, we expect nuclear structure effects to be definitely larger than in CEvNS and presumably at least at the 10$\%$ level. To properly assess uncertainties in these observables,  modern machine learning techniques will be used~\cite{Furnstahl:2014xsa, Co:2020gwl}.


\section{Connecting Electron- and Neutrino-Nucleus Scattering Physics}
\label{sec:elec-to-neutrino}


In this section, we briefly describe how the underlying nuclear physics, probed by electrons and neutrinos, is intimately connected to each other. The benefit of kinematically controlled beams and precise measurements makes the electron scattering data a vital benchmark to constrain nuclear models intended to be used in neutrino experiments. 


\subsection{Vector and Axial Current}
\label{subsec:vector_axial}
The fact that electron and (anti)neutrino scattering processes both originate from the same electroweak interaction 
opens opportunities to invest synergies between both, and exploit similarities between these two types of processes to
gather information about the weak interaction that is much more difficult to assess experimentally than its electromagnetic counterpart is.
Whereas electromagnectic interactions are in first-order mediated by a (virtual) photon, weak processes are effectuated by 
 the exchange of a $W^{\pm}$ or $Z^0$ boson for charged and neutral
 weak processes, respectively.
However, the way electromagnetic and weak processes manifest themselves is often very different. 
This is not in the least due to the fact
that intrinisic strength of the weak interaction is extremely small compared to that of the electromagnetic one.  On top of that,
the massless photons lead to a $Q^{-4}$ dependence of the boson propagator and hence cross sections that are strongly forward peaked in the direction
of the outgoing lepton, whereas for neutrinos transverse contributions and backward lepton scattering plays a more important role over a broader
 kinematic range.

Electromagnetic interactions on the nucleon or nucleus, mediated by the neutral photon, do not change charge, and consist of  isoscalar and third component 
isovector contributions.  
This combination is shared by neutral weak interactions, but charge-changing weak processes on the other hand  are of a pure 
isovector nature. 
As a  consequence of vector current conservation (CVC), the vector form factors in weak scattering, notoriously difficult to investigate experimentally,
can be inferred by a mere rotation in isospin space i.e.~CVC provides an unambiguous
relation between the vector form factors for both types of interactions. In a similar way,  the structure of the hadronic response functions 
for the vector contributions is fully equivalent for both the electromagetic and weak processes.  For the axial current
a softer constraint is provided by the partial conservation of the axial current (PCAC). 
As a consequence of the fact that the chiral symmetry of the Langragian is here only
approximate, the axial current conservation  becomes exact only in the limit of vanishing pion mass. The relative weight of vector and axial
contributions in neutrino-induced processes depends strongly on the kinematics of the interaction.  

In semi-leptonic processes,  the leptonic probe, be it neutral or charged, is of course always subject to weak interactions. Obviously, in the weak case the interference between vector
and axialvector contributions  gives rise to parity violating processes. 
In  electron scattering processes the axial contribution 
is however veiled by the much stronger electromagnetic vector contributions.  In parity violating electron scattering (PVES) the interference between electromagnetic and weak
processes i.e. the interference between processes in which a photon or a Z boson are exchanged, dominates the process, thereby providing 
a way to amplify the axial contribution by multiplying it  with the much larger
vector electromagnetic term.  PVES and experiments carried out with polarized electrons hence provides a powerful tool to 
study axial characteristics of the weak interaction in experiments with 
charged leptons as probes, that allow for a better control of the kinematics.

\subsection{Nuclear Effects}
From the nuclear point of view, the influence of nuclear medium effects as e.g.~Pauli blocking, long- and short-range correlation, and medium modifications of resonances at higher energies can be expected to be largely the same for electron as for neutrino-induced processes. An important point of attention is of course the charge-changing nature of charged current weak processes.  In CC quasi-elastic processes, the final nuclear states will be members of the same isospin triplet as the initial one,
 but reside in a different final nucleus. 
As a consequence, the effect of the changed nuclear (and Coulomb) potential experienced by hadronic and leptonic ejectiles in charged-current weak interactions
has to be taken into account carefully. 
  Owing to the vector-axial and parity-violating character
of the interaction, weak processes are able
to excite different nuclear states than electromagnetic processes do, a feature that is particularly apparent for low-energy processes. 
So can neutrinos e.g.~induce $0^{-}$ transitions in nuclei, which electromagnetic interactions cannot, thus opening a window to study nuclear states that are difficult to 
reach otherwise. The effects of two-body currents in nuclei, both in the vector and the axial sectors, are comparable to other nuclear effects. Two-body currents often are inseparable to the one-body current even in the one-particle emission channel and present an additional challenge~\cite{Martinez-Consentino:2021vcs}. Likewise, at much higher energies, pion production processes for neutrinos can lead to different final states and outgoing hadrons than their 
electron-induced counterparts and will require hence more elaborate modeling efforts with less experimental constraints available.

Whereas validating modeling efforts for neutrino processes by also applying the models to electron scattering data 
 is a general  practice and essential for a solid verification of the vector part of the model, some models go to lengths to  exploit the similarities between both interaction types to 
guide the description of weak scattering processes.  
Standing out in this respect is the SuSA model \cite{Gonzalez-Jimenez:2014eqa} that found its origin exactly in the analogies between neutrino- and electron
 quasi-elastic processes,
and relies on scaling considerations as a strong tool to predict neutrino-induced cross sections.

\subsection{CEvNS and PVES} 
\label{subsec:cenns_pves}


CEvNS and PVES are intimately connected to each other.
From the formal point of view, both processes are described in first order perturbation theory via the exchange of an electroweak gauge boson between a lepton and a nucleus. While in CEvNS the lepton is a neutrino and a $Z^0$ boson is exchanged, in PVES the lepton is an electron, but measuring the asymmetry allows one to select the interference between the $\gamma$ and $Z^0$ exchange. As a result, both the CEvNS cross section (see Section~\ref{sevens}) and the PVES asymmetry (see Section~\ref{section6}) depend on the weak form factor $F_W(Q^2)$, which  is mostly determined by the neutron distribution within the nucleus.  The latter builds an even stronger anchor between CEvNS and PVES.

Parity violating electron scattering experiments offer the least model dependent and most precise approach to experimentally probing the neutron distribution. Any result that will come from the PVES program with the goal of pinning down the neutron-skin thickness will help improving our understanding of the weak form factor and hence influence  CEvNS.
However, CEvNS  has also been proposed as an alternative
and attractive opportunity in the future to constrain the neutron distribution and the neutron radius in nuclei~\cite{Amanik:2009zz,Patton:2012jr,Cadeddu:2017etk}, provided that enough statistics can be reached. 

The main difference lies in the choice of the nuclear target, which is determined by practical considerations. In the case of PVES, the targets need to be stable (or almost stable) neutron-rich nuclei, such as $^{208}$Pb and $^{48}$Ca, that do not present low-lying excited states that would contribute to the background noise. 
In the case of CEvNS, isotopes of sodium, argon, germanium,  cesium and iodine will be used, as the low cost allows to build large detectors with these materials.

 Because various electroweak observable correlate~\cite{Yang:2019pbx} with each other,
theoretical calculations will  help to further connecting the various nuclear targets and the two endeavors of CEvNS and PVES.  For example, we can expect that constraints  experimentally determined on the neutron-skin thickness of one nuclear target will affect the prediction of the weak form factor of another target (see also Section~\ref{sevens}). 

It is worth mentioning that another promising avenue in experimental extractions of the neutron-skin thickness is measurements of coherent pion photoproduction cross sections~\cite{Tarbert:2013jze,Thiel:2019tkm}. Experiments with $^{58}\mathrm{Ni}$, $^{116,120,124}\mathrm{Sn}$, and $^{208}\mathrm{Pb}$ were performed last decade~\cite{Tarbert:2013jze,Bondy:2015uoa}. Both experimental and theoretical developments are expected for this process. Similar to PVES, measurements on a few isotopes can help to remove major theoretical uncertainties.

\subsection{Experimental Input}
 

Electron and neutrino scattering are strongly linked in theory because electron interactions with hadrons are almost purely vector and neutrinos interact through a mixture of vector and axial currents.  However, there are significant differences in the measurements that have been published so far because electron beams are more intense and monoenergetic.  Therefore, both quality and depth of measurements are more complete and neutrino calculations should take advantage of this as much as possible.

The vector electron interaction is closely linked to the vector neutrino interaction through CVC (Sect.~\ref{subsec:vector_axial}). There are various ways to accomplish this union.  Theorists tend to write separate codes for electron and neutrino interactions consistent with the underlying symmetries.  Event generators tend to build the relationships into a single code.  For example, GENIE~\cite{Andreopoulos:2009rq,GENIE:2021npt} uses the relationship given in Quigg's book~\cite{Quigg:2013ufa} to transform neutrino interactions into electron interactions.  

The ideal implementation of electron interactions in event generators should allow adjustment of the models to electron scattering data simultaneous with the update of neutrino interactions.  Vector form factors in neutrino models should come from same code as electron models and the core interaction strengths linked by theory.  Neutral current (NC) neutrino interactions are closely related to electron scattering, but charged current (CC) neutrino interactions can be different. For example, the excitation of nucleon resonances goes to different charge states for NC and CC neutrino interactions but the decay of the resonances should be universal.

Although GENIE has a generally solid implementation of electron scattering, it suffers from the fact that the code was originally designed for neutrino interactions and electron interactions were added later.  Recent changes~\cite{electronsforneutrinos:2020tbf} have brought GENIE closer to the ideal but problems in implementation still cause troubles with the excitation of nucleon resonances. Implementation of electron scattering in NuWro and NEUT is still underway. Further discussion on the importance of electron scattering data for improving the neutrino event generators can be found in Ref.~\cite{Campbell:2022qmc}. Significant effort will be required to reach the proper level so that the knowledge gained is implemented in neutrino event generators with the proper parameters and their associated uncertainties.  To accomplish this, a more complete set of electron semi-inclusive (esp. $(e,e'\pi)$) data are needed.


\section{Experimental Landscape I: Input to Accelerator-Based Neutrino Program}
\label{sec:expt_landscape_acc-based}



For over five decades, the electron scattering experiments at different facilities around the world have provided wealth of information on the complexity of nuclear structure, dynamics and reaction mechanisms. Decades of experimental work has provided vital testbed to assess and validate theoretical approximations and predictions that propelled the theoretical progress staged around~\cite{Benhar:2006wy, Benhar:2006er}. While previous and existing electron scattering experiments provide important information, new measurements which expand kinematic reach, final states and provide an updated estimate of uncertainties are beneficial.

In this section, we briefly outline current and planned electron scattering experiments, majority of them are motivated by the needs of the accelerator neutrino experiments. For each experiment, we briefly lay out the experimental setup, capabilities and describe their (performed or proposed) measurements. We then attempt to identify connections and gaps between these experimental programs in the context of their input to the accelerator neutrino program.

\subsection{Current and Planned Experiments}



\subsubsection{E12-14-012 at JLab}\label{subsubsec:E12-14-012}

\subsubsection*{Overview}
The JLab E12-14-012 experiment, performed in Jefferson Lab Hall A, has collected inclusive $(e,e^\prime)$ and exclusive $(e, e^\prime p)$ electron-scattering data in five different kinematical setups using a natural argon and a natural titanium targets. 

Under the basic assumption that scattering processes involve just individual nucleons, and not considering FSI between the outgoing proton and the spectator nucleons, the removal energy and momentum of the knocked out particle, $E$ and ${\bf p}$, could be reconstructed from measured kinematical variables, and the cross section can be written in simple factorized form in terms of the spectral function of the target nucleus, $P({\bf p},E)$, trivially related to the nucleon Green's function, $G({\bf p},E)$, through $P({\bf p},E) =  \frac{1}{\pi}\  {\rm Im} \ G({\bf p},E)$~\cite{Benhar:2015wva}.
As a consequence, the spectral function which represents the probability to remove a proton with momentum ${\bf p}$ from the target nucleus leaving the residual system with excitation energy $E-E_{\rm thr}$, with $E_{\rm thr}$ being the proton emission threshold---is determined from the data. For nuclei as heavy as $^{40}$Ar, FSI effects should be accurately taken into account using an 
optical potential~\cite{Giusti:1987eoi}.

Over the past several years, a great deal of work has been devoted to applying the spectral function formalism to the study of neutrino-nucleus interactions, whose quantitative understanding is needed for the interpretation of accelerator-based searches of neutrino oscillations, see, e.g., Refs.~\cite{Benhar:2015wva,Ankowski:2016jdd}.

\subsubsection*{Experimental Setup}
The experiment E12-14-012 was performed at Jefferson lab in Spring 2017. Inclusive $(e,e^\prime)$ and exclusive $(e,e^\prime p)$ electron scattering data were collected on targets of natural argon and natural titanium, the average neutron numbers calculated according to the natural abundances of isotopes are 21.98 for argon and 25.92 for titanium~\cite{Murphy:2019wed}. 

The E12-14-012 experiment used an electron beam of energy 2.222~GeV provided by the Continuous Electron Beam Accelerator Facility (CEBAF) at Jefferson Lab. The average beam current was approximately 15~$\mu$A for the $^{40}$Ar target and 20~$\mu$A for the $^{48}$Ti target. The scattered electrons were momentum analyzed and detected in the left high-resolution spectrometer (HRS) in Hall A and the coincident protons were similarly analyzed in the right HRS. 

The spectrometers are equipped with two vertical drift chambers (VDCs) providing tracking information~\cite{Fissum:2001st}, two scintillator planes for timing measurements and triggering, double-layered lead-glass calorimeter, a gas Cherenkov counter used for particle identification~\cite{Alcorn:2004sb}, pre-shower and shower detectors (proton arm only)~\cite{Alcorn:2004sb} and pion rejectors (electron arm only)~\cite{Alcorn:2004sb}. The HRSs were positioned with the electron arm at central scattering angle $\theta_e=21.5^{\circ}$ and the proton arm at an angle $\theta_{p^\prime}=-50{^\circ}$. The beam current and position, the latter being critical for the electron-vertex reconstruction and momentum calculation, were monitored by resonant radio-frequency cavities (beam current monitors~\cite{Alcorn:2004sb}) and beam position monitors~\cite{Alcorn:2004sb}). The beam size was monitored using harp scanners.

The experiment employed an aluminum target with a thickness of $0.889\pm0.002$~g/cm$^2$. One of the aluminum foils was positioned to match the entrance and the other to match the exit windows of the argon gas target cell. The two thick foils were separated by a distance of 25~cm, corresponding to the length of the argon gas cell and the Al foil's thickness.

Table~\ref{tab:kinematic} summarizes the kinematical conditions of the E12-14-012 experiment.

\begin{table}[tb]
\centering
\caption{\label{tab:kinematic}Kinematics settings used to collect the data for the JLab E12-14-012 experiment.}
\begin{tabular}{ c | c c c c c c c c c c }
\hline
    & $E_{e}^\prime$    & $\theta_e$& $Q^2$         & $|P|$     & $T_{p^\prime}$& $\theta_{p^\prime}$   & $|{\bf q}|$ & $p_m$ & $E_m$ \\
    & (GeV)             & (deg)     & (GeV$^2/c^2$) & (MeV/$c$) & (MeV) & (deg)    &(MeV/$c$)     & (MeV/$c$) & (MeV) \\
\hline
    kin1 & 1.777        & 21.5      & 0.549        & 915         & 372  & $-50.0$   & 865          &   50 		& 73  \\
    kin2 & 1.716        & 20.0      & 0.460        & 1030        & 455   & $-44.0$  & 846          &   184 		& 50  \\
    kin3 & 1.799        & 17.5      & 0.370        & 915         & 372  & $-47.0$   & 741          &   174		& 50  \\
    kin4 & 1.799        & 15.5      & 0.291        & 915         & 372  &  $-44.5$  & 685          &   230      & 50  \\
    kin5 & 1.716        & 15.5      & 0.277         & 1030       & 455  &  $-39.0$  & 730          &   300      & 50  \\
\hline
\end{tabular}
\end{table} 

\par The VDCs' tracking information was used to determine the momentum and to reconstruct the direction (in-plane and out-of-plane angles) of the scattered electron and proton, and to reconstruct the interaction vertex at the target. Both the electron and proton arm information was separately used to reconstruct the interaction vertex and we found them in very good agreement. The transformation between focal plane and target quantities was computed using an optical matrix, the accuracy of which was verified using the carbon multi-foil target data and sieve measurements as described in previous papers~\cite{Dai:2018gch,Dai:2018xhi,Murphy:2019wed}. Possible variations of the optics and magnetic field in both HRSs are included in the analysis as systematic uncertainties related to the optics.

Several different components were used to build the triggers: the scintillator planes on both the electron and proton spectrometers, along with  signals from the gas Cherenkov (GC) detector, the pion rejector (PR), the pre-shower and the shower detector (PS). 

\subsubsection*{Measurements}
\begin{figure}
\centering
\includegraphics[width=0.49\columnwidth]{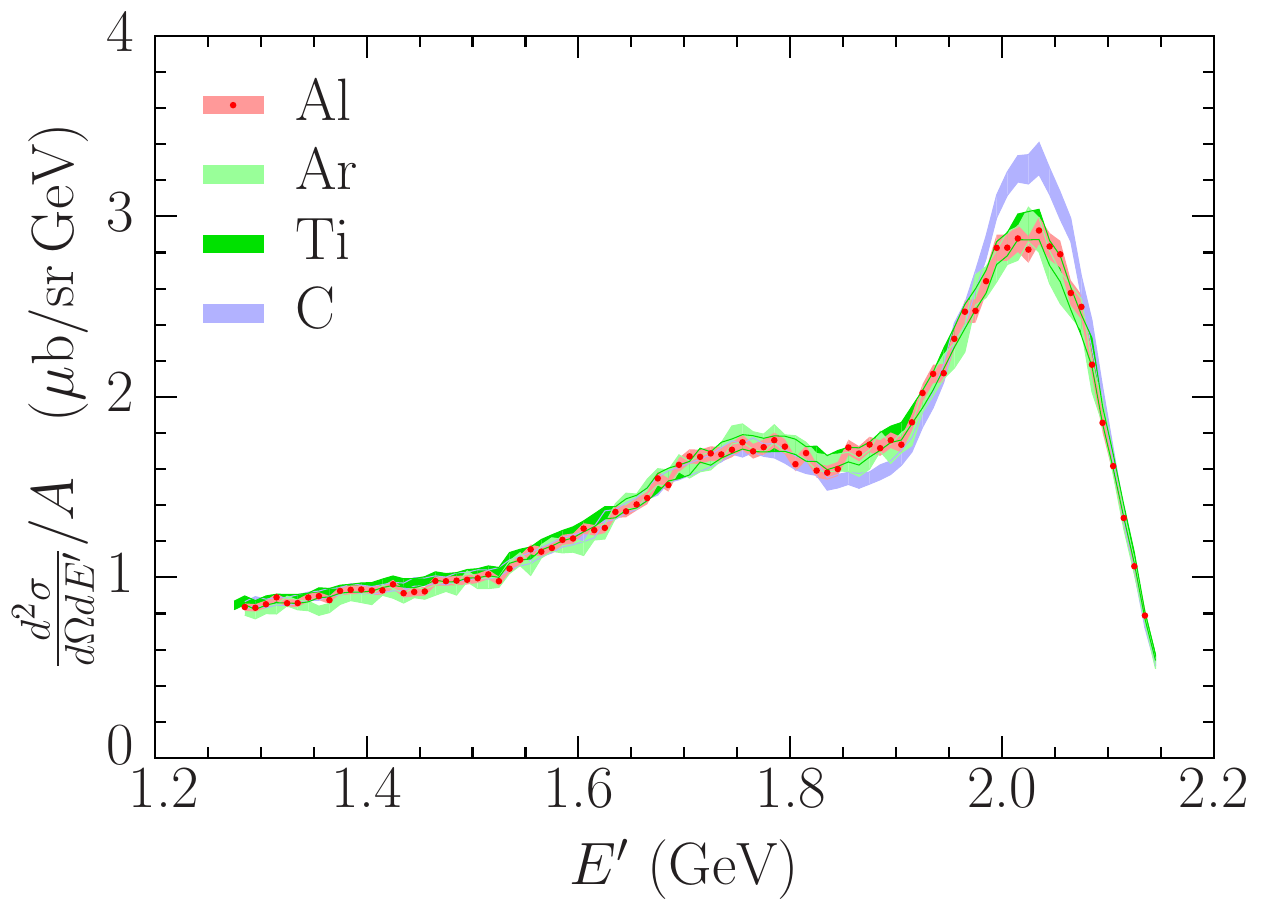}
\includegraphics[width=0.48\columnwidth]{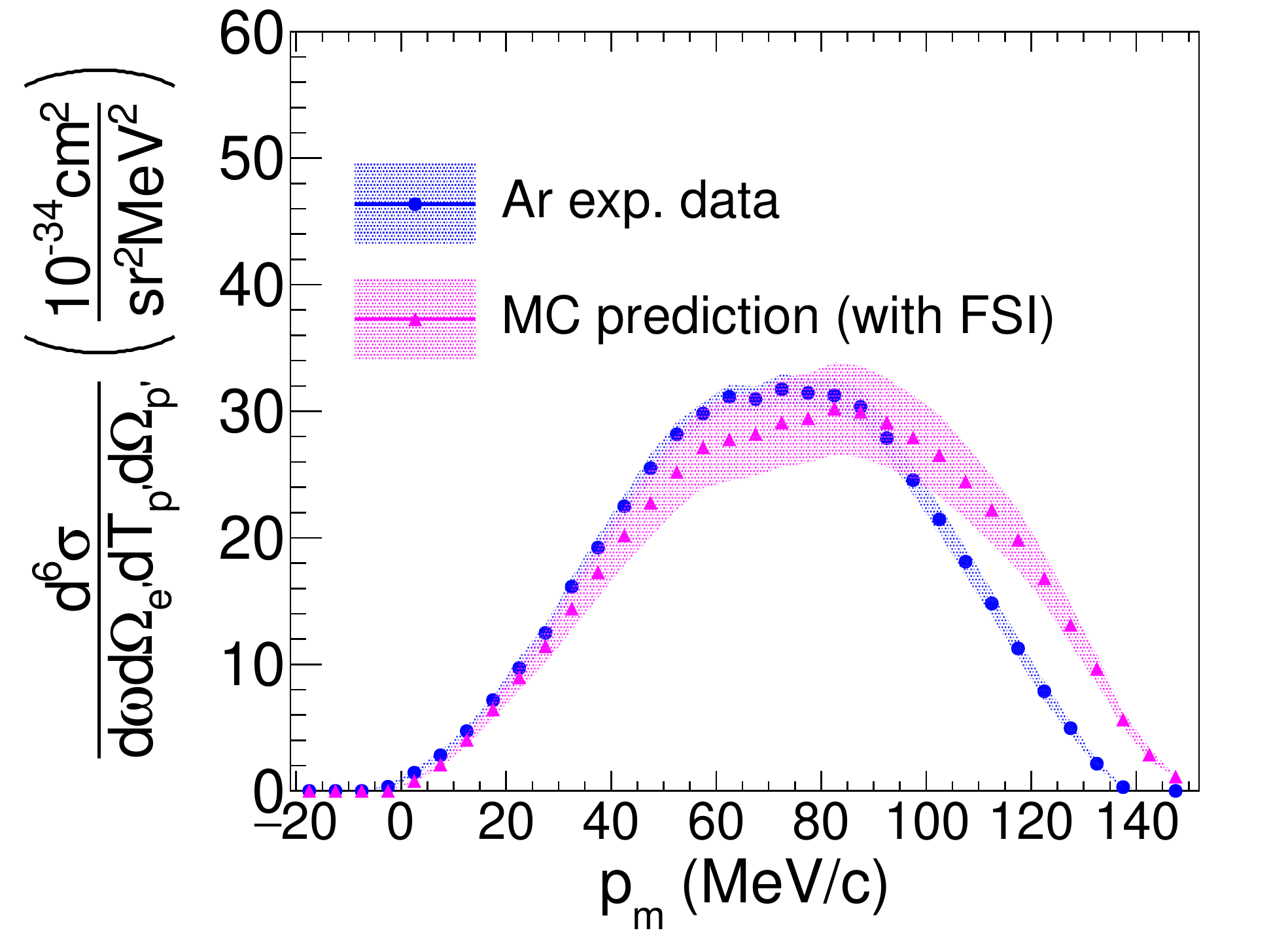}
\caption{(Left) Comparison of the cross sections per nucleon for aluminum, argon~\cite{Dai:2018gch}, titanium~\cite{Dai:2018xhi}, and carbon~\cite{Dai:2018xhi} measured at beam energy 2.222~GeV and scattering angle $15.54^\circ$. The average nucleon number for every target is calculated according to the natural abundances of isotopes, see details in the text. The bands represent the total uncertainties. Figure from Ref.~\cite{Murphy:2019wed}. (Right) Six-fold differential cross section for argon as a function of missing momentum integrated over $E_m$ from 0 to 27~MeV. Figure from Ref.~\cite{JeffersonLabHallA:2020rcp}.}
\label{fig:E12-14-012}
\end{figure}

The inclusive data analysis is summarized in Refs.~\cite{Dai:2018xhi,Dai:2018gch,Murphy:2019wed} and presented in Fig.~\ref{fig:E12-14-012} (left), while the exclusive data analysis for Ar is summarized in Refs.~\cite{JeffersonLabHallA:2020rcp} and represented in Fig.~\ref{fig:E12-14-012} (right).

%
%

The inclusive cross section has been measured on a variety of targets, including aluminum, carbon, and titanium and a closed argon-gas cell. The data extend over broad range of energy transfer, where quasielastic interaction, $\Delta$-resonance excitation, and deep inelastic scattering yield contributions to the cross section. The double differential cross sections are reported with high precision ($\sim$3\%) for all targets over the covered kinematic range (\cite{Dai:2018xhi,Dai:2018gch,Murphy:2019wed}).

The exclusive cross section has been measured on Ar with $\approx$4\% uncertainty, and have been studied as a function of missing energy and missing momentum, and compared to the results of Monte Carlo simulations, obtained from a model based on the Distorted Wave Impulse Approximation~\cite{JeffersonLabHallA:2020rcp}.

The availability of the inclusive data will be essential for the development of accurate models of the lepton scattering off argon and titanium~\cite{Barbieri:2019ual}. The ultimate goal of the exclusive analysis is the determination of the spectral functions. The results of this effort will provide the input for sampling momentum and 
energy of the target nucleons, thus allowing for a reliable reconstruction of neutrino interactions in liquid argon detectors. 
 

\subsubsection{E4nu at JLab}
\label{subsubsec:E4nu}

\subsubsection*{Overview}
The $e4\nu$ effort at JLab uses a large acceptance detector to measure  wide phase-space exclusive and semi-exclusive electron-nucleus scattering  at known beam energies to test energy reconstruction methods and interaction models.  This includes both existing data taken with the CLAS spectrometer in 1999  and new data taken in winter 2021/22 with the CLAS12 spectrometer.

The incident electron energies of 1 to 6 GeV span the range of typical accelerator-based neutrino beams, the nuclear targets are similar, and the hadron detection thresholds are similar to those of tracking neutrino detectors. The data analysis is as similar as possible to neutrino-based ones~\cite{MicroBooNE:2020fxd}.  For each  exclusive channel,  differential cross sections are extracted using experimental neutrino  methods. The electron–nucleon cross-section is much more forward peaked than the neutrino cross-section due to the difference in the  photon and $W^\pm$ propagators. Therefore, each event is weighted by $Q^4$, where $Q^2$ is the square of the four-momentum  transfer.
The cross sections, typically as a function of the reconstructed lepton energy, are compared to predictions from the electron mode of neutrino-nucleus interaction event generators and to the true beam energy (see Fig.~\ref{fig:e4nuerec}). 

\subsubsection*{Experimental Setup}
Data was taken in 1999  using the CEBAF large acceptance spectrometer (CLAS) at the Thomas Jefferson National Accelerator Facility (JLab)  to measure  electron-nucleus scattering with multi-hadron final states.  Data was taken with $^3$He, $^4$He, C and Fe targets at 1.1, 2.2 and 4.4 GeV.  CLAS used a toroidal magnetic field with six sectors of drift chambers, scintillation counters, Cherenkov counters and electromagnetic calorimeters to identify electrons, pions, protons, and photons, and to reconstruct their trajectories\cite{CLAS:2003umf}. CLAS covered scattering angles from about $10^\circ$ to about $140^\circ$ with azimuthal coverage of 50 to 80\%.  CLAS covered approximately 50\% of $4\pi$ and had reconstruction thresholds of approximately 150 MeV/c for charged pions and 300 MeV/c for protons.  Photons of greater than 250 MeV were detected at $8\le\theta\le 45^\circ$. The experiment used the standard inclusive electron trigger, with Cherenkov Counters (CC) and Electromagnetic Calorimeter (EC) at 1.1 and 2.2 GeV and EC only at 4.4 GeV. The experiment  collected data corresponding to  $2.2\times 10^9$ triggers and 13~mC of beam charge. 

More data was taken with CLAS12 in 11/2021--2/2022 as described below.

\subsubsection*{Modelling Development}
In order to compare the $e4\nu$ measurement to models, the electron-scattering mode of the GENIE event generator (e-GENIE) was revised~\cite{electronsforneutrinos:2020tbf}. Many of the interaction models were debugged and updated, making e-GENIE self-consistent with the neutrino codes for the first time~\cite{GENIE:2021npt}. Radiative effects necessary for comparing QE-like electron scattering data to simulation was incorporated. The simplistic implementation 
was vetted against electron-hydrogen scattering data from JLab. In addition, models based on the SuSAv2 approach were implemented for the QE and MEC processes. 

Neutrino mode cross sections are very similar to electron-mode cross sections scaled by $Q^4$, validating the use of electron scattering to improve neutrino event generators. 
The e-GENIE predictions were compared with inclusive electron scattering data, and agree moderately  well at low energy transfer.  However there is significant  disagreement at higher energy transfer, which is dominated by resonance production~\cite{electronsforneutrinos:2020tbf}. 

\subsubsection*{Measurements}
The first  $e4\nu$ analysis used 1999 CLAS \cite{CLAS:2003umf}
data  on $^{4}$He, C and Fe targets with  monochromatic electron beams of 1.159, 2.257 and 4.453 GeV. The analysis focused on Quasi Elastic (QE) like events, by selecting events with one electron, one proton of momentum larger than 300 MeV/c, and no charged pions of momenta larger than 150 MeV/c. In addition, the minimum CLAS electron scattering angle corresponded to minimum momentum transfers of $Q^2> 0.1, 0.4,$ and  $0.8$ GeV$^2$ for beam energies of 1.159, 2.257, and 4.453 GeV, respectively. The QE-like event sample was corrected for events with  undetected pions, photons, and extra protons using a data-based method; more details about the background subtraction can be found at\cite{CLAS:2021neh}.  

This was the first test of incident lepton reconstruction. Two methods  were used to reconstruct the incident energy: (1)  using only the scattered lepton momentum and angle and assuming QE kinematics: \[E_{QE}=\frac{2M_N\epsilon+2M_NE_l-m_l^2}{2(M_N-E_l+k_l\cos\theta_l)} .
\] This  method is used in  experiments with Cherenkov-detectors, which are sensitive only to the outgoing lepton~\cite{T2K:2019bcf}. (2)  used the outgoing proton and electron energies and is commonly used across tracking detectors:
\[E_{cal}=E'_e+T_p+E_{binding}.
\]
Cross sections as a function of  reconstructed energies were extracted  for C and Fe at 1.159, 2.257, 4.453 GeV and compared to two different versions of GENIE~\cite{CLAS:2021neh}.  

Fig.~\ref{fig:e4nuerec} from \cite{CLAS:2021neh} shows the cross section as a function of the calorimetric reconstructed energy.  All spectra show a sharp peak at the real beam energy, followed by a large tail at lower energies. For carbon, only 30–40\% of the events reconstruct to within 5\% of the real beam energy. For iron this fraction is only 20–25\%, highlighting the crucial need to model precisely the low-energy tail of these distributions. GENIE described the data qualitatively, but there are significant quantitative disagreements.  One version of GENIE predicted a reconstructed energy that was about 20 MeV too low.  This disagreement in the tails increases with incident  energy and nuclear mass. 

The data was then binned as a function of  the transverse missing momentum, \[P_T=P^e_T+P^p_T
\]
(where $P_T^e$ and $P_T^p$ are the three-momenta of the detected lepton and proton perpendicular to the direction of the incident lepton). At  $P_T\le 200$ MeV/c almost all events reconstruct to the correct cross section and GENIE described those events well.  At intermediate $P_T$ ($200 \le P_T\le 400$ MeV/c) only a small fraction of events reconstruct to the correct beam energy and GENIE estimates the cross section of events that reconstruct to lower energies.  At large $P_T$ ($P_T>400$ MeV/c), none of the events reconstruct to the correct beam energy and GENIE significantly overestimates the  cross section of events that reconstruct to lower energies.  This shows that neutrino experiments can use $P_T$ to select events that reconstruct to the correct energy and that GENIE requires further modeling improvements for large $P_T$~\cite{CLAS:2021neh}.

A second analysis examining the $1p1\pi$ channel is in progress for this data set.  There is very little data for this channel in the resonance region for nuclear targets and the largest disagreement between GENIE and inclusive $A(e,e')$ data~\cite{electronsforneutrinos:2020tbf} is seen there. These data will help improve all generator models for nuclear targets.  

\subsubsection*{New Data with CLAS12}

A dedicated $e4\nu$  measurement with the upgraded CLAS12 detector  was approved by the JLab PAC with an A scientific rating for 45 beam days~\cite{e4vproposal}. 
The experiment will  use targets from D to Sn, including neutrino-detector materials (C, O, and Ar), at 1, 2, 4, and 6  GeV. During the fall/winter 2021/2022 data were collected with   C and Ar at 2, 4 and 6 GeV, along with measurements of H, D, $^{4}$He, $^{40}$Ca, $^{48}$Ca and Sn for calibration and for other nuclear physics purposes.

The  CLAS12 spectrometer operates at higher luminosity (ten times higher than CLAS), and takes advantage of the lower scattering angle thresholds ($\theta_e\ge 8^\circ$), along with the neutron detection capabilities in the central detector ($\approx 35^\circ \le \theta\le 135^\circ$) \cite{Burkert:2020akg} .  This experiment will provide a much larger data set, with more event channels ($1p0\pi$, $1p1\pi$, etc), greater angular and kinematic coverage, and more targets than the existing CLAS data. 

\subsubsection{LDMX at SLAC}\label{subsubsec:LDMX}

LDMX (Light Dark Matter eXperiment) is a fixed-target electron-scattering experiment planned to search for sub-GeV dark matter~\cite{LDMX:2018cma}. A signature expected for dark-matter production is missing energy and missing momentum, carried away from the event by undetected particle(s). To reach the parameter space previously unexplored, LDMX needs to achieve an excellent background rejection and analyze an extensive body of data. 

The requirement of precise reconstruction of both charged and neutral hadrons, in conjunction with the necessity to collect a vast number of electron-scattering events makes the goals of LDMX highly synergistic with the physics program of the neutrino community interested in improving current modeling of electroweak interactions~\cite{Ankowski:2019mfd}.

Dedicated studies of neutrino cross sections  performed by the MINERvA Collaboration at the kinematics similar to that of DUNE unambiguously show that existing Monte Carlo generators are not able to explain available data~\cite{MINERvA:2014rdw,MINERvA:2019kfr,MINERvA:2021wjs}. Particularly dire problems are observed for pion production,  
where measurements for different channels within a~single experiment cannot be reconciled, and the origin of this issue is difficult to pinpoint~\cite{MINERvA:2019kfr}. This is a consequence of the complexity of the problem: flux-averaging greatly diminishes differences between different interaction channels at the inclusive level, different interaction mechanisms yield the same final states, only nuclear cross sections are available with the desired precision, and scattering involves multicomponent vector and axial currents~\cite{Ankowski:2020qbe}.

These difficulties can be addressed by making the full use of the similarities of electron and neutrino interactions~\cite{Ankowski:2019mfd}. Electron-scattering measurements feature monoenergetic and adjustable beams, cross sections higher by several orders of magnitude, and feasible targets ranging from complex nuclei to deuterium and hydrogen. As long as the cross sections for neutrinos and electrons are calculated consistently, and they are (over)constrained by precise electron data, the uncertainties of Monte Carlo simulations employed in neutrino physics can be reduced to those stemming from the axial contributions, to be constrained by the near-detector measurements. In the context of an accurate calorimetric neutrino-energy reconstruction, the exclusive cross sections---currently unavailable for electron scattering on the targets of interest---play particularly fundamental role~\cite{Friedland:2018vry,Friedland:2020cdp}.

\begin{figure}
    \begin{center}
        \includegraphics[width=0.46\textwidth]{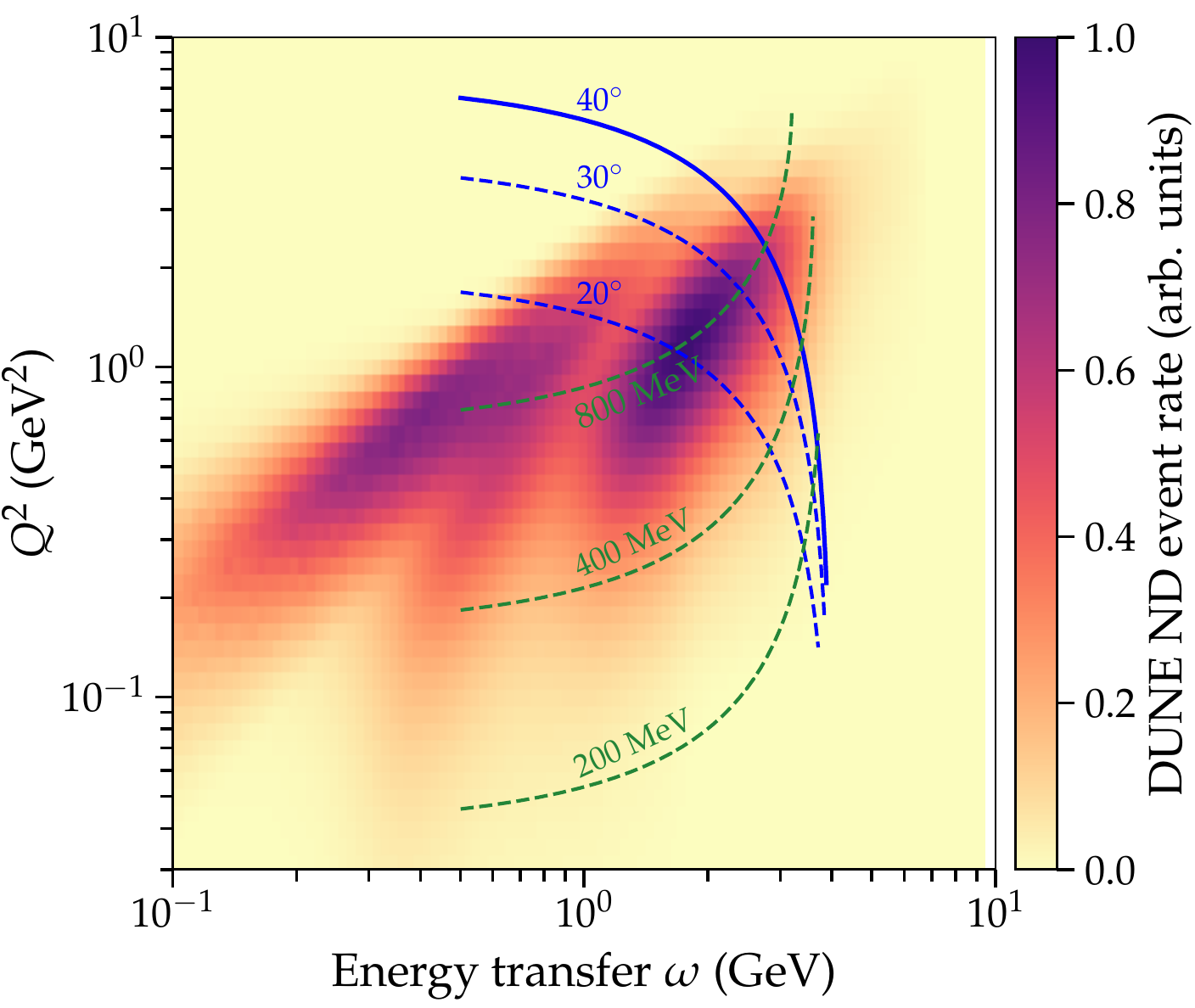}
        \hspace{0.05\textwidth}
        \includegraphics[width=0.46\textwidth, height = 7.05cm]{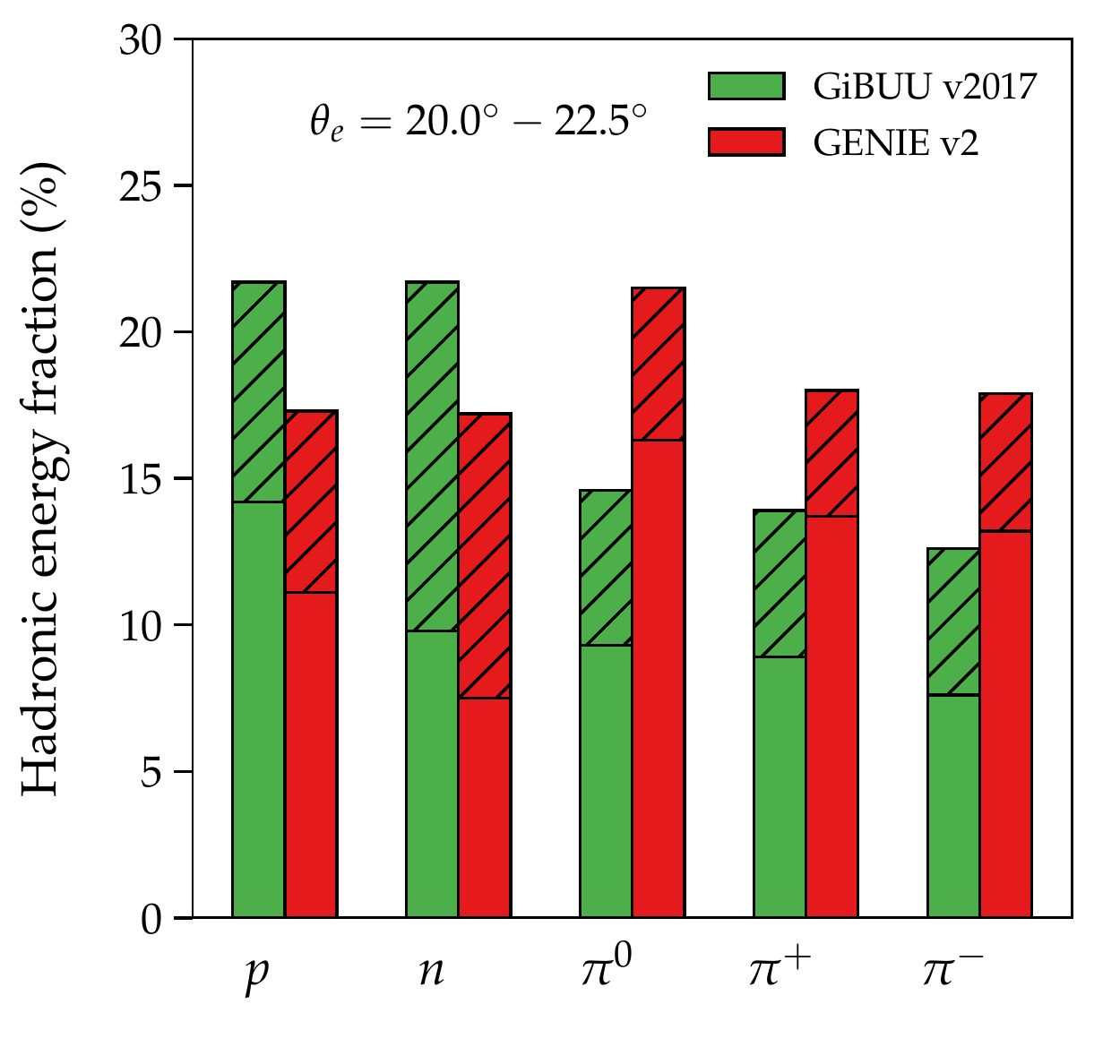}
       \caption{(Left) Kinematics accessible in \emph{inclusive and (semi)exclusive} electron-scattering measurements at LDMX with a 4-GeV beam, overlaid on the charged-current $\nu_\mu$Ar event distribution expected in the DUNE near detector according to GiBUU~\cite{Ankowski:2019mfd}. Constant scattering angles $\theta_e$ and transverse momenta $p^e_T$ of electrons correspond to the blue and green lines, respectively. LDMX will probe the region with $\theta_e <40^\circ$ and $p^e_T >10$~MeV, extending beyond the plot. (Right) Breakup of hadronic energy of events with $20.0^\circ\leq\theta_e\leq22.5^\circ$, $p^e_T >200$~MeV, and $\omega > 1$ GeV, according to GiBUU and GENIE. The contributions outside of the LDMX acceptance---outside the 40$^\circ$ cone or below detection thresholds---are hatched~\cite{Ankowski:2019mfd}.}
        \label{fig:LDMX_coverage}
    \end{center}
\end{figure}

{\it Inclusive measurements.} Currently, only one kinematics of relevance for DUNE has been probed through inclusive electron scattering on argon~\cite{Dai:2018gch} and titanium~\cite{JeffersonLabHallA:2018zyx}, the proton structure of which mirrors the neutron structure of argon. The data---collected for beam energy 2.22~GeV, scattering angle $15.54^\circ$, and extending to energy transfers below 0.95 GeV---show that at the inclusive level and within uncertainties, the cross sections per nucleon do not differ between argon and titanium~\cite{Murphy:2019wed}, see Fig.~\ref{fig:E12-14-012}. Employing a~titanium target~\cite{Ankowski:2019mfd}, LDMX can collect a wealth of inclusive data for beam energy 4-GeV, scattering angles below $40^\circ$, and energy transfers above 1 GeV. To illustrate their importance for DUNE, in the left panel of Fig.~\ref{fig:LDMX_coverage}, the kinematics accessible in LDMX is overlaid on the distribution of charged-current $\nu_\mu$ interactions with argon expected in the near detector of DUNE, as calculated using GiBUU~\cite{Ankowski:2019mfd}. With its nominal selections, LDMX can cover the deep-inelastic kinematics in great detail, and probe the tail of resonance production. These cross sections, currently unavailable, would be essential for validation of Monte Carlo simulations for DUNE~\cite{Ankowski:2020qbe}. 

{\it (Semi)exclusive measurements.} With full-event reconstruction being essential for its core program and the envisioned large integrated luminosity, LDMX is uniquely equipped to extract coincidence cross sections for a variety of final states involving pions and nucleons. For example, of the $10^{14}$ electrons on target envisioned for the first phase of low-luminosity running, $10^8$ will scatter at $20.0^\circ$--$22.5^\circ$ and pass the selection cuts. The kinetic energy spectra are expected to be measured down to $\sim$60 MeV for charged pions and protons, and to $\sim$500 MeV for neutrons. As summarized in the right panel of Fig.~\ref{fig:LDMX_coverage}, the collected data should allow discriminating between the cross section models implemented in GiBUU and GENIE, performing their extensive validations, and quantifying a number of uncertainties originating from the rich physics of electroweak interactions.

{\it Future potential.} By employing titanium as the target,
LDMX in the baseline dark-matter configuration can perform measurements of both inclusive and (semi)exclusive electron scattering, which are of great interest for DUNE. The physics program can be further extended, as discussed in Ref.~\cite{Ankowski:2019mfd}.
\begin{itemize}
\item{Varying the target material, LDMX would provide more data
for nuclear modeling, allowing for deeper understanding
of the cross-sections’ dependence on the
atomic number. In particular, a high-pressure argon-gas cell would directly address the needs of the neutrino community, complementing the titanium results. Measurements for helium, deuterium, and hydrogen would provide a
handle on how nuclear transparency affects the
exclusive cross sections. They would not only allow a~clean separation of hadronic
and nuclear effects, but also give an unprecedented insight into how the effects of final-state interactions shape the observed hadronic spectra.}

\item{The nominal physics selections can be extended to energy transfers below 1~GeV, to fully cover the regions in which resonance production and meson-exchange
currents provide important contributions to the cross sections. Such measurements could improve our understanding of resonant and nonresonant pion production, and shed light on the composition of hadronic final states in the dip region between the quasielastic and $\Delta$-production peaks.}
\end{itemize}


\subsubsection{A1 Collaboration at MAMI}
\label{subsec:A1_MAMI}


The Mainz Microtron (MAMI) is an electron accelerator run by the Institute for Nuclear Physics of the University of Mainz and it employed for nuclear hadron physics experiments. MAMI is a multi-stage racetrack microtron with normal conducting acceleration cavities. The accelerator is available for experiments since 1991 and has continuously undergone further development. The last added stage, MAMI-C, became operational in 2007 rising the beam energy up to 1.6 GeV with a maximum current of 100 $\mu$A. The beam diameter is $\sim$0.1~mm with an energy resolution of 13 keV. A beam stabilization system maintains the beam position constant within 200$\mu m$. The beam has a continuous-wave structure (100\% duty cycle) and the possibility to be polarized at the $\sim 80\%$ level. The main parameters of the last two MAMI microtron stages are summarized in Tab.~\ref{tab:accel}. 

\begin{table}
\centering
{\small
\begin{tabular}{l|c|c}
\hline
{\bf Stage}                   & {\bf MAMI B}      & {\bf MAMI C} \\
\hline
{\bf Max. Energy}             & 855.1 MeV   & 1508 MeV \\
{\bf Circulations}	        &    90       & 43 \\
{\bf Deflecting Magnets:}     &             & \\
\quad Magnetic Field          & 1.28 T      &	0.95 - 1.53 T \\
\quad Mass                    & 2$\times$450 t & 4$\times$250 t \\
{\bf Cavities:}               &             & \\
\quad Frequency               & 2.45 GHz    & 2.45 / 4.90 GHz \\
\quad Power	                & 102 kW      & 117 / 128 kW \\
{\bf Linac Length}            & 8.9 m	      & 8.6 / 10.1 m \\
\hline
\end{tabular}
}
\caption{Main properties of the MAMI cascaded microtron accelerator.}
\label{tab:accel}
\end{table}

The MAMI beam can be delivered in different experimental halls. The A1 Collaboration hall in particular
is devoted to experiments with electrons, while the A2 collaboration hall converts the electrons into
a tagged photon beam. Another hall is devoted to experiments with X-rays and detector development.
The A1 collaboration operates a three magnetic spectrometer setup for experiments in nuclar and hadron physics. The spectrometers are conventionally called A, B, and C. Spectrometers A and C have the same quadrupole-sextupole-dipole-dipole (QSDD) magnetic optics for maximizing angular acceptance, while spectrometer B has only one dipole magnet: this construction choice allows this spectrometer to reach smaller angles (i.e. to move closer to the beamline) with the additional capability to move out-of-plane. 

\begin{table}
\centering
{\small
\begin{tabular}{l|c|c|c|c}
\hline	                  &   {\bf units}   &   {\bf Spec. A}    &   {\bf Spec. B}   &   {\bf Spec. C} \\
\hline
{\bf Magnetic Configuration}	  &    -      &    QSDD	     &      D      &    QSDD \\
{\bf Maximum momentum}	  &  (MeV/c)  & 735          &   870       &  551 \\
{\bf Reference momentum}	  &  (MeV/c)  & 630          &   810       &  459 \\
{\bf Central Momentum}	  &  (MeV/c)  & 665          &   810       &  490 \\
{\bf Solid Angle Acceptance}	  &    (msr)  &  28          &   5.6       &   28 \\
{\bf minimum angle}	          &    (deg)  &	 18          &     7       &    8 \\ 
{\bf maximum angle}		  &    (deg)  & 160 	     &    62       &  160 \\
{\bf Momentum acceptance}	  &     --    &	20\%         &    15\%     &   25\%\\
{\bf Angular acceptance:}        &           &              &             &      \\
\quad dispersive	          &   (mrad)  &  70          &    70       &   70\\
\quad nondispersive             &   (mrad)  & 100          &    20       &  100\\
{\bf momentum resolution}       &     --    & 10$^{-4}$     & 10$^{-4}$   &  10$^{-4}$ \\    
{\bf angular resolution at target}  & (mrad)&	$<3$         &    $<3$     & $<3$\\
{\bf position resolution at target} &	(mm)  &	3--5	     &    1        & 3--5\\
\hline
\end{tabular}
}
\caption{Main properties of the three magnetic spectrometers of the A1 Collaboration at MAMI.}
\label{tab:specs}
\end{table}

The spectrometers have a momentum resolution of $10^{-4}$. The characteristics of the spectrometers
are summarized in Tab.~\ref{tab:specs}. All spectrometers have the same detector package composed of two vertical drift chambers (VDCs), two planes of scintillator bars, and a Cherenkov detector. Each VDC is composed by 4 wire planes, two perpendicular to the dispersive plane for momentum and out-of-plane angle reconstruction, and the other 2 planes, with wires tilted by a few degrees for completing the position determination in the focal plane. The single plane efficiency is $\sim$99\%, leading to an overall efficiency of better than 99.9\%. The scintillator planes serve as trigger detectors and for particle identification, together with the Cherenkov detector. The spectrometers can be employed in coincidence (double or triple) for exclusive measurements. Different kinds of targets can be installed at the end of the MAMI beamline at the pivot point around which the spectrometers can be rotated. Solid-state targets (e.g. C, Ca, Si, Ta, Pb)
are routinely used. A cryogenic target is also available and has been in use with different elements in
liquid phase (H, $^2$H, $^3$He, $^4$He) and its use for noble gases (e.g. Ar, Xe) is possible. A waterfall target is also available for measurements with oxygen. Recently, a supersonic gas-jet target was successfully tested with hydrogen and argon.


\subsubsection{A1 Collaboration at Spanish facilities}

\subsubsection*{Experimental setup and capabilities}

The proposed instrument ({\bf eALBA}) is a multipurpose electron beam facility, where the electron beam from the ALBA synchrotron in Barcelona(Spain) is extracted from the tunnel to the experimental hall, where a multipurpose experimental area will be conditioned. This beam is expected to serve different experimental applications: Detectors characterization for HEP, Gamma detector characterization,  Radiation hardness tests for space electronic,  Electron imaging, Microdosimetry and Radio-biology Experiments with conventional and Ultra-high Dose electron, but also Education and Outreach. This infrastructure will be unique in Europe covering  wide beam energy range (100~MeV - 3~GeV) with excellent repetition rate (3~Hz) and high intensity per bunch (1~nC).

One of the proposed experimental facilities is devoted to Electron-Nucleus scattering experiments for neutrino studies. One possible realization of this experiment is an atmospheric pressure Time Projection Chamber (TPC) surrounding a target located outside of the gas volume.  The TPC operates in a magnetic field pointing in the direction of the electron beam and the TPC field.  The magnetic field bends particles according to its transverse momentum.  The field strength is expected to be moderate, and it can be built using conventional technology. A cylindrical TPC with a diameter of 1 m and a length 2 m is technologically feasible and sufficient for the purpose of the experiment.  The inner wall of the TPC can be very thin when operating the TPC at atmospheric pressure and allowing low momentum particles to enter in the active TPC volume from the external beam target. There are available technology to read TPC of this size using Micro Pattern Gas detector technology as the one pioneered by the T2K near detector. By using existing technology and a moderate magnetic field, the experiment cost can be reduced to levels that be constructed by small collaborations. 
 
\begin{figure}
\centering
\includegraphics[width=0.49\columnwidth]{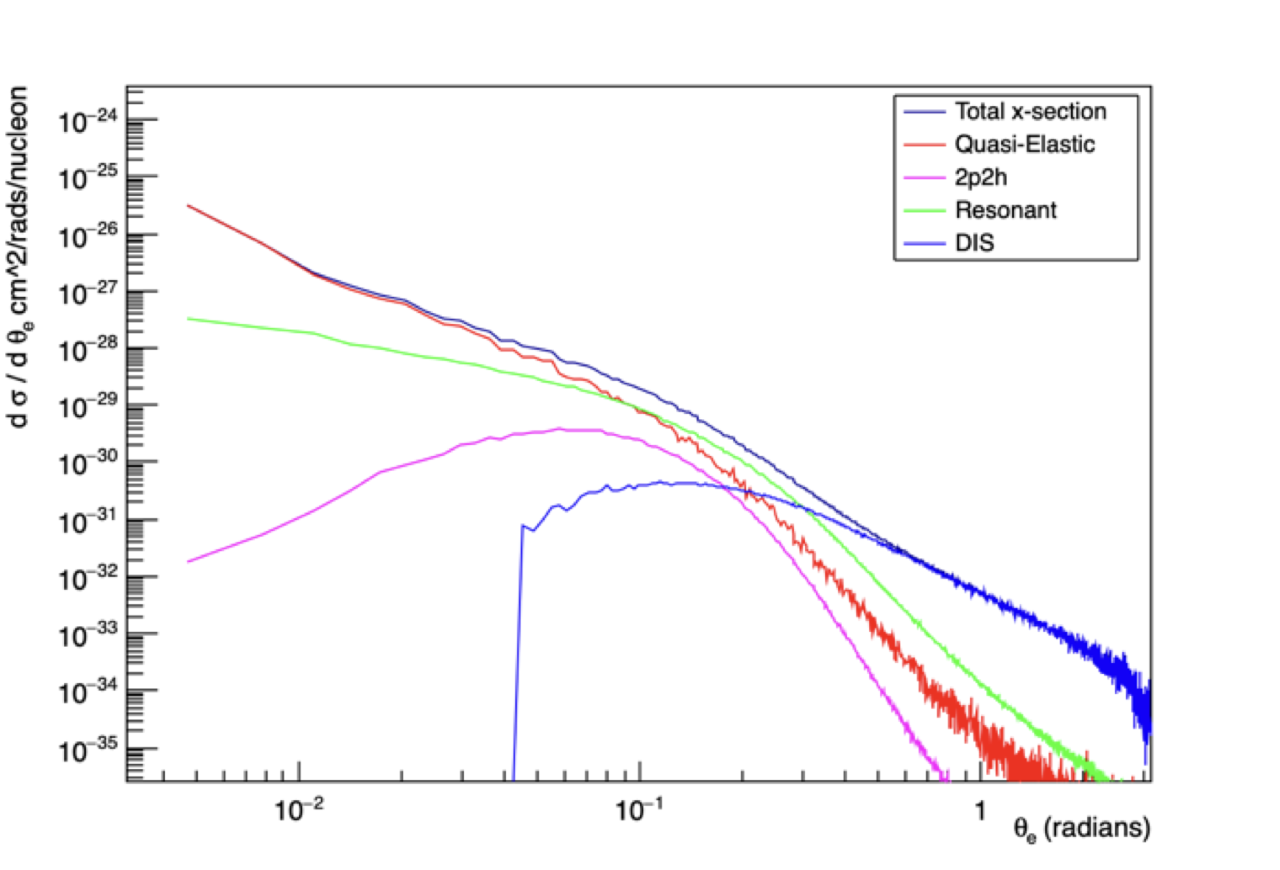}
\caption{Angular distribution of 4~GeV electrons scattered from the C nucleus as predicted by the GIBUU event generator}
\label{fig:ealba}
\end{figure}

The angular distribution of 4~GeV electrons scattered from the C nucleus as predicted by the GIBUU event generator 
are shown in Figure~\ref{fig:ealba}.
Requesting the electrons to traverse at least 25 cm inside the TPC (a fourth of the total length) with an inner TPC radius of 1cm, the minimum acceptance angle is around $40$ mrads. Most of the interactions below this value will be quasielastic and probably will not eject hadrons with a large transverse momentum. Electrons ejected below $40$ mrads can be detected with a forward calorimeter.

One of the drawbacks of TPC’s is the low drift velocity in the gas, of the order of tenths of microseconds. According to the same GIBUU simulations, the integrated cross-section above $4$ mrads is of the order of $2 \times 10^{-33} \mbox{cm}^2/\mbox{nucleon}$. The interaction probability in a Carbon target with a density of approximately $2 g/$cm$^3$ is of the order of $5\times10^{-7}$ in a $1$ mm thick target. At a $1 pA$ beam current, there will be $\approx 3$ interactions per second with the scattered electron above $40$ mrads. This event rates are within the TPC timing capabilities. Most of the interactions below $40$ mrads will leave no signature in the detector due to the spiral trajectory caused by its low transverse momentum.  Targets thinner than 1mm will improve the performance, the lower interaction rate can be compensated with the high beam intensities and the target will affect less the hadrons produced in the interactions. A dedicated Monte Carlo study will be carried out to understand more precisely the capabilities of this design.  

By hosting an external target, the experiment can explore different nuclei targets and energies. To reach levels interesting to neutrino experiments based on the water Cherenkov technology, we will require electron beams with energies from 500 MeV to few GeV.   

\subsubsection*{Proposed measurements}

The measurements will cover several target nuclei. We expect a minimum set of targets: C, CH, Be, Ca. Gaseous and liquid targets (He, H$_2$0, Ar,...) are a challenge due to the size of the containment vessel. The experiment can be seen as a facility that could be run in the future serving the needs of different experiments. 

\subsection{Summary of the ongoing and future efforts}

\begin{table}
\centering
{\small
\begin{tabular}{ l | c c c c }
\hline {\bf Collaborations} & {\bf Kinematics} & {\bf Targets} & {\bf Scattering} \\
\hline
{\bf E12-14-012 (JLab)} & $E_e$ = 2.222 GeV & Ar, Ti & ($e,e'$) \\
{(Data collected: 2017)} & $15.5^\circ\leq{\theta_e}\leq21.5^\circ$ & Al, C & $e, p$ \\
 &$-50.0^\circ\leq{\theta_p}\leq-39.0^\circ$ &  & in the final state \\
\hline
{\bf e4nu/CLAS (JLab)} & $E_e$ = 1, 2, 4, 6 GeV& H, D, He, & ($e,e'$) \\
{(Data collected: 1999, 2022)} & ${\theta_e} > 5^\circ$  & C, Ar, $^{40}$Ca, & $e, p, n, \pi,\gamma$ \\
 &  & $^{48}$Ca, Fe, Sn & in the final state \\
\hline
{\bf LDMX (SLAC)} & $E_e$ = 4.0, 8.0 GeV & & ($e,e'$) \\
  {(Planned)} & ${\theta_e} < 40^\circ$ & W, Ti, Al & $e, p, n, \pi, \gamma$ \\
 &  &  &  in the final state \\
\hline
{\bf A1 (MAMI)} & 50 MeV $\leq E_e \leq 1.5$ GeV & H, D, He & ($e,e'$) \\
 {(Data collected: 2020)} & $7^\circ\leq{\theta_e} \leq 160^\circ$ & C, O, Al & 2 additional \\
 {(More data planned)} &  & Ca, Ar, Xe & charged particles & \\
\hline
{\bf A1 (eALBA)} & $E_e$ = 500 MeV & C, CH & ($e,e'$) \\
 {(Planned)} & ~~~~~~- few GeV & Be, Ca & \\
\hline
\end{tabular}
}
\caption{Current and planned electron scattering experiments.}
\label{tab:e-scattering_landscape}
\end{table}

\begin{figure}
    \begin{center}
        \includegraphics[width=0.99\textwidth]{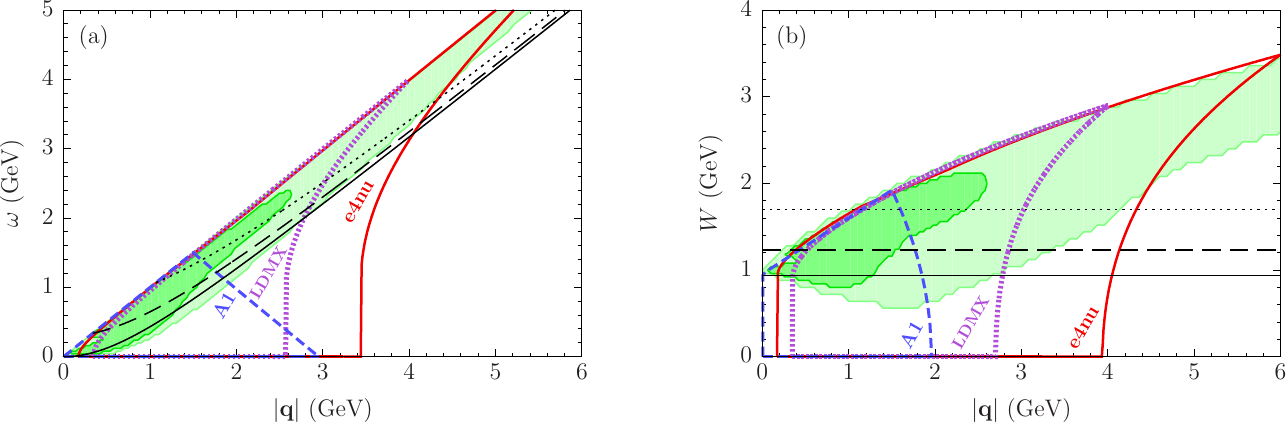}
       \caption{Kinematic coverage of the ongoing and planned experiments for inclusive and exclusive electron scattering on targets including argon and titanium, presented in the (a) $(|\mathbf{q}|, \omega)$ and (b) $|\mathbf{q}|, W)$ planes, with momentum transfer $\mathbf{q}$, energy transfer $\omega$, and hadronic mass $W$. The light and dark shaded areas cover 68\% and 95\% of charged-current $\nu_\mu$Ar events expected in the DUNE near detector~\cite{DUNE:2021cuw}, according to GENIE 3.0.6. The thin solid, dashed, and dotted lines correspond to the kinematics of quasielastic scattering, $\Delta$ excitation, and the onset of deep-inelastic scattering at $W = 1.7$ GeV on free nucleons.}
        \label{fig:future_coverage}
    \end{center}
\end{figure}


In Table~\ref{tab:e-scattering_landscape}, we present a summary of the current and planned electron-scattering experiments, which cover a broad range of kinematics and carry varied level of particle identification and other detection capabilities. This kinematics is presented in Fig.~\ref{fig:future_coverage}, where it is overlaid on the regions expected to contain 68\% and 95\% of charged current interactions of muon neutrinos with argon in the DUNE near detector, as estimated using GENIE 3.0.6. Thanks to its broad coverage of scattering angles---from $7^\circ$ to $160^\circ$---and beam energies---from $\sim$50 MeV to 1.5 GeV---A1 in MAMI is uniquely equipped to perform extensive studies of the quasielastic  and $\Delta$ peaks. LDMX in SLAC is planned to feature a~4-GeV beam, closely corresponding to the average energy in DUNE, and perform extensive studies of pion production. The e4nu experiment in JLab has the potential to study a large swath of the DUNE kinematics, thanks to the broad range of employed beam energies. To probe the kinematic region of interest at higher momentum transfers and perform studies in the deep-inelastic regime, an experiment at even higher energies would be desirable, such as the 8-GeV configuration of LDMX (not depicted). However, the issue deserving most attention should always remain measuring  exclusive cross sections, essential for accurate calorimetric reconstruction of neutrino energy, especially for low-energy hadrons.



\section{Experimental Landscape II: Input to Low-Energy Neutrino Physics}
\label{sec:expt_landscape_low-energy}

The tens-of-MeV neutrinos, from stopped pion sources or from core-collapse supernova, primarily interact via two processes: coherent elastic neutrino-nucleus scattering (CEvNS), and inelastic charged and neutral current scattering. Precise determination of Standard Model CEvNS cross section will enable new physics searches in CEvNS experiments, while precise inelastic cross section determination will enable detection of supernova signals in DUNE experiment.

The main source of uncertainty in the evaluation of the CEvNS cross section is driven by the underlying nuclear structure, embedded in the weak form factor, of the target nucleus.  Weak form factors and neutron radii are historically poorly known. Recent advancements in PVES experiments, utilizing polarized electron beams, provide relatively model-independent ways of determining weak form factors that can be used as direct input in determining CEvNS cross section. The inelastic neutrino-nucleus cross sections in this tens-of-MeV regime are quite poorly understood. There are very few existing measurements, none at better than the 10\% uncertainty level. As a result, the uncertainties on the theoretical calculations of, e.g., neutrino-argon cross sections are not well quantified at all at these energies. To this end, 10s of MeV electron scattering data will be vital in constraining neutrino-nucleus interaction at this energy scale.  

Dedicated electron scattering experiments with targets and kinematics of interests to low-energy neutrino experiments will be crucial in achieving precision goals of low-energy neutrino programs. 

\subsection{Parity-Violating Electron Scattering Experiments}
\label{section6}
\subsubsection{PREX and CREX at JLab}

The parity violating asymmetry $A_{pv}$ for elastic electron scattering is the fractional difference in cross section for positive helicity and negative helicity electrons.  In Born approximation $A_{pv}$ is proportional to the weak form factor $F_W(q^2)$ \cite{Horowitz:1999fk},
\begin{equation}
    A_{pv}=\frac{d\sigma/d\Omega_+-d\sigma/d\Omega_-}{d\sigma/d\Omega_++d\sigma/d\Omega_-}=\frac{G_Fq^2|Q_W|}{4\pi\alpha\sqrt{2}Z}\frac{F_W(q^2)}{F_{ch}(q^2)}\, .
    \label{eq.Apv}
\end{equation}
Here $F_{ch}(q^2)$ is the (E+M) charge form factor that is typically known from unpolarized electron scattering.  Therefore, one can extract $F_W$ from measurements of $A_{pv}$.  Note that Eq. \ref{eq.Apv} must be corrected for Coulomb distortions as discussed in ref. \cite{Horowitz:1998vv}.  These effects are absent for neutrino scattering.

The PREX experiment at JLAB measures $A_{pv}$ for $\approx 1$ GeV electrons elasticly scattered from $^{208}$Pb \cite{PREX:2021umo}.  This experiment was motivated both to aid the interpretation of atomic parity experiments and to study neutron rich matter in astrophysics.


\paragraph{Atomic parity nonconservation (PNC):} provides an important low energy test of the standard model.  The weak interaction involves the overlap of atomic electrons with the weak charge of the nucleus (ie neutrons).  The existing atomic Cs experiment measures the weak charge to about 0.3\% and is probably not limited by uncertainty in the neutron radius.  However, a future 0.1\% atomic experiment may need $R_n$ to about 1\%. The best PNC test of the standard model may involve combining an atomic experiment with a separate measurement of $F_W(q^2)$ \cite{Horowitz:1999fk}.

\paragraph{Neutron rich matter in astrophysics:} Compress almost anything enough and electrons capture on protons to make neutron rich matter.  This material is at the heart of many fundamental questions in Physics and Astrophysics.  What are the high density phases of QCD? Where did chemical elements come from? What is the structure of many compact and energetic objects in the heavens, and what determines their electromagnetic, neutrino, and gravitational-wave radiations? 

A heavy nucleus such as $^{208}$Pb has a skin of neutron rich matter \cite{Thiel:2019tkm}.  Probing this region with coherent neutrino scattering or parity violating electron scattering allows the study of neutron rich matter in the laboratory .  The pressure of neutron rich matter pushes neutrons out against surface tension in $^{208}$Pb.  This same pressure holds neutrons up against gravity in a neutron star.  A neutron star is 18 orders of magnitude larger than a nucleus but it is made of the same neutrons, with the same strong interactions, and the same equation of state or pressure as a function of density.  Therefore measurement of the neutron radius of $^{208}$Pb in the laboratory has important implications for the structure of neutron stars \cite{Fattoyev:2017jql,Reed:2021nqk}.



Table \ref{tab:parityviolation} lists three recent parity violating elastic electron scattering experiments.  PREX has determined the neutron radius $R_n$ of $^{208}$Pb to 1.3\% \cite{PREX2}.  The CREX experiment has measured $A_{pv}$ from $^{48}$Ca \cite{CREX}.  The analysis of this measurement is ongoing but we expect a sensitivity to $R_n$ in $^{48}$Ca of $\pm$0.7\%. Finally the Qweak collaboration has measured electron scattering from $^{27}$Al ~\cite{QWeak:2021ivq}.  Note because of the limited energy resolution of the Qweak spectrometer $\pm 100$ MeV this experiment is limited by systematic errors from inelastic contributions rather than from statistics.  

We see that the best present parity violating experiments have determined $R_n$ to around one \%. This may be more accurate than present neutrino experiments. Although, the PVES experiments are usually carried out at a single value of the momentum transfer at a time.  Nevertheless, one may be able to use parity violating electron scattering (and nuclear theory) to constrain nuclear structure and $F_W(q^2)$.  This could allow coherent neutrino scattering experiments to better constrain nonstandard neutrino interactions.

\begin{table}[tb]
\centering
\begin{tabular}{ c | c c c c }
 Experiment    & Target& $q^2$ (GeV$^2$)  & $A_{pv}$ (ppm)     & $\pm\delta R_n$ (\%) \\
\hline
PREX & $^{208}$Pb & 0.00616 & $0.550\pm0.018$ & 1.3 \\
CREX & $^{48}$Ca & 0.0297 & & 0.7\\
Qweak  & $^{27}$Al & 0.0236 & $2.16\pm 0.19$ & 4\\
MREX  & $^{208}$Pb & 0.0073 & & 0.52\\
\hline
\end{tabular}
\caption{\label{tab:parityviolation}Parity violating elastic electron scattering experiments.}
\end{table}

\subsubsection{MREX at MESA}
\label{subsec:mrex}

A novel experimental program is foreseen at the future "Mainz Energy-recovering Superconducting Accelerator" (MESA), which is being build at the Institute of Nuclear Physics of the Johannes Gutenberg University (Mainz)\cite{1923127}. MESA is a low-energy high-intensity electron accelerator for particle physics at the precision and intensity frontier. The combination of high beam quality and intensity allows precision measurements to explore the limits of the standard model.  The design of the MESA accelerator provides two modes of operation, which serve as the basis for a totally different classes of experiments. With a “Pseudo internal target” combined with two high resolution spectrometers experiment using ERL-mode explorative searches for hitherto unknown gauge bosons can be undertaken (see \ref{subsec:magix}) 

The second experiment (P2 \cite{Becker:2018ggl}) is a parity violating electron scattering experiment with spin-polarized beam using EB-mode. The aim of the experiment P2 is the precise determination of a fundamental parameter of the Standard Model: the effective electroweak mixing angle $\sin^2(\theta_W)$. The desired relative accuracy of 0.1\% outperforms existing measurements at low-energy scales by more than an order of magnitude and is thus comparable to the measurements at LEP and SLAC from the 1990s at very high-energy scales. The aimed sensitivity is similar to the best Z-pole measurement.

Preliminary feasibility studies have shown that the Mainz Radius EXperiment (MREX) at MESA, by adopting the same detector setup intended for the P2 experiment will be able to extract the neutron skin thickness of $^{208}$Pb by a determination of the neutron radius with a sensitivity of $\delta R_\text{n}/R_\text{n}=\SI{0.5}{\%}$ (or $\pm$ 0.03 fm). 
With the necessary momentum resolution of 1\%, MREX will thus outperform the PREX experiment. Indeed although the maximum incident energy at MESA (thus the cross section) will be lower compared to the JLab measurement, the experiment will benefit on the one hand from the higher beam intensities and on the other hand from the full azimuthal coverage of the detector. 
A peak sensitivity for $^{208}$Pb is expected for the parameter reported in table \ref{tab:overview}.
Since the same experimental setup will be used to determine the weak mixing angle $\theta_{W}$ and the neutron skin of $^{208}$Pb the running time for the latter one was restricted to about 1500 hours. 

\begin{table}
  \centering
  \begin{tabular}{cc}
	\toprule[1.5pt]
	beam energy & $\SI{155}{MeV}$ \\
	beam current & $\SI{150}{\mu A}$ \\
	target density & $\SI{0.28}{g/cm^2}$ \\
	polar angle step size & $\Delta\theta=\ang{4}$ \\
	polar angular range & \ang{30} to \ang{34} \\
	degree of polarization & \SI{85}{\%} \\
	parity violating asymmetry & \SI{0.66}{ppm}\\
	running time & 1440 hours \\
	\hline
	systematic uncertainty & \SI{1}{\%}\\
	$\delta A^\text{PV}/A^\text{PV}$ & \SI{1.39}{\%} \\
	$\delta R_\text{n}/R_\text{n}$ & \SI{0.52}{\%}\\
  \bottomrule[1.5pt]
  \end{tabular}
  \caption{Kinematical values and general parameters used for the run time estimate.}
  \label{tab:overview}
\end{table}

The ultimate precision on the neutron radius measurement at MESA will be limited by the implicit model assumption in the method on the surface thickness. The extraction of the neutron-skin thickness in parity violation experiments is indeed based on the measurement of the weak form factor at a single value of momentum transfer. Thus some assumptions concerning the surface thickness have to be made in order to reliably extract the neutron-skin thickness.  In order to remove this model dependence a determination of the surface thickness is necessary with a precision better than the current model assumption (\SI{25}{\%})\cite{Reed:2020fdf}. In particular since the parity-violation asymmetry measurement at MESA will determine the neutron radius of $^{208}$Pb with a precision of $\pm$\SI{0.03}{fm} the surface thickness must be known down to \SI{10}{\%}. Preliminary sensitivity study indicates that such a measurement can be performed at the A1 spectrometer at MAMI (see \ref{subsec:A1_MAMI}). According to the construction of the spectrometers and their arrangement on the pivot surrounding the scattering chamber, there are limitations in the accessible angular range. Furthermore, only selected beam energies at MAMI are equipped with a special stabilization system which is essential for performing parity-violation experiments. Based on the experience of the first successful experimental campaigns \cite{Esser:2018vdp} and on the requirement that the sensitivity of $A_{\mathrm{PV}}$ to the weak-charge radius must be suppressed w.r.t. the sensitivity to the surface thickness for the given beam energy and scattering angles the optimum kinematics is obtained for 570 MeV electron energy and \SI{0.06}{GeV^2/c^2} and \SI{0.02}{GeV^2/c^2} momentum transfer for the two spectrometers respectively.

\subsubsection{Identifying Connections and Gaps}

The true breakthrough in our understanding of neutron-rich matter will be only attained by combining results from both terrestrial laboratories and astrophysical observations. The ultimate precision on the neutron-skin thickness will be attained at the MESA accelerator. However already nowadays from the comparison between the most recent PREX-II result and the astrophysical observation from both gravitational wave and neutron star observatories stronger constraints on the EOS of neutron-rich matter have been determined \cite{Reed:2021nqk}.

On the other side a recent analysis of the combination of coherent elastic neutrino-nucleus scattering data from the COHERENT experiment and atomic parity violation data has determined the most precise measurement of the neutron rms radius and neutron-skin values
of $^{133}$Cs  and $^{127}$I  \cite{Cadeddu:2021ijh}. A combination of complementary methods to extract the neutron radius will naturally produce even stronger constraints.

As already pointed out in \ref{subsec:cenns_pves} CEvNS experiments are limited to detector materials with low scintillation or ionisation thresholds in order to efficiently measure low-energy nuclear recoils.
Quite the contrary is needed as target material in parity violation electron scattering experiments: in this case the highest the excited state of the nucleus the lower the contamination of the elastic asymmetries by inelastic contributions from the excited state (as an example the first excited state in $^{208}$Pb  is about 2.60 MeV). In addition due to the high intensity of the electron beam a high melting temperature of the target material is as well desirable.

A direct comparison between the neutron radius extracted for the same nucleus in both type of experiments is therefore technically not feasible in the near future.  Naturally a combination of these two different approaches can proceed by theory. In particular parity violation electron scattering results can be used as anchors for modern density functional theory calculations which can then predict the coherent nuclear form factor as measured in CEvNS at low momentum transfer.
Ultimately CEvNS extraction of the neutron distribution for heavy nuclei by benchmarking theoretical predictions can serve as additional constraint to the EOS of neutron rich matter.

The measurement of the coherent nuclear form factor for medium-mass nuclei as for example $^{40}$Ar could additionally provide a validation bridge for modern ab-initio calculations based on effective field potential.
Together with the neutron density extraction for $^{48}$Ca and $^{27}$Al as determined by CREX and Qweak these data will provide the most accurate benchmark of modern nuclear structure calculations.

\subsection{Low Energy Electron Scattering}

\subsubsection{MAGIX Collaboration at MESA}
\label{subsec:magix}


As discussed in Sec.~\ref{subsec:mrex}, MESA, a new cw multi-turn energy recovery linac for precision particle and nuclear physics experiments with a beam energy range of 100-200 MeV is currently being built. MESA  will be operated in two modes: energy recovery mode (ERM) and external beam mode (XBM). In ERM, the accelerator will provide a beam current of up to 1 mA at 105 MeV for the MAGIX internal target experiment with multi-turn energy recovery capability. In XBM, a polarized beam of 150 $\mu$A will be provided. The linac will have an energy gain of 50 MeV/pass by using four ELBE-like 9-cell cavities installed in two cryomodules. The MAGIX experiment will be installed on one of the recirculating paths: in this way the beam will pass through a supersonic jet-target and then re-injected in the accelerator for energy recovery. MAGIX will employ two high-resolution magnetic spectrometers movable around the jet target. The relatively low density of the target (areal thickness $\sim 10^{20}$ atoms/cm$^{2}$/Z$^2$ with Z the gas atomic number) will be compensated by the large ($\sim$mA) beam current.

The two spectrometers will be used for the detection of scattered electrons and other produced particles. They will be equipped with TPCs (time projection chambers) with GEM readout for tracking, and with a trigger and veto system consisting of a stack of plastic scintillation detectors planes and a variable number of additional lead absorber layers.
Silicon strip detectors will be installed inside the scattering chamber for detecting protons and nuclear fragments in coincidence with the spectrometers.

The MESA accelerator and the MAGIX setup are currently under construction and the jet target was already built and tested at the University of Münster. 
In 2017 the target was successfully installed and operated with hydrogen at the existing A1 three-spectrometer setup at the Mainz Microtron (MAMI) electron accelerator in Mainz, Germany. In 2021, the target was successfully tested with argon. The jet target will be operated with hydrogen, oxygen, and noble gases (e.g. He, Ar, Xe). Currently, the opportunity to install a beam-dump is investigated: this would allow also the use of solid-state targets (which are too dense to allow the beam recirculation) like carbon, for example.

MAGIX was conceived with the main goals of measuring the nucleon form factor at low $Q^2$ contributing to the "proton-radius puzzle", searching for light dark sector particles, measure astrophysical S-factors, and precisely measure exclusive and inclusive nuclear cross sections. The MESA relatively low beam energy falls into the supernova neutrinos range and can thus be interesting for the tuning and testing of neutrino generators in this energy range.


\section{Addressing NP and HEP Boundary Conditions}
\label{sec:np-hep}





A ongoing challenge of the work described in this white paper-- adapting electron scattering to neutrino physics-- is the separation of HEP and NP research communities and a lack of funding support to address these cross-cut research topics. Theorists and experimentalists who work on electron scattering are supported from NP, but oscillation experiments like DUNE are supported by HEP. To avoid duplication of funding, funding agencies opt to avoid cross cutting topics such as this work; this view was expressed by multiple groups via a  workshop survey. Further, as some of the experiments (like DUNE) are DOE HEP projects, the work is viewed as therefore under only HEP or DOE scope.

Unfortunately, this separation has historically harmed the science and community members. For example, when MINERvA was founded, collaborators funded from NP had trouble getting funding for the (HEP) MINERvA project, and had to leave. This resulted in the loss of valuable cross field discussion. It is difficult to train and retain early career scientists with expertise on intersectional topics, as they do not `fit' neatly into either category. This separation does not seem to occur in the LHC, where there is a clear recognition of the value of cross section calculation and generator effort. 

We propose the following possible solutions to challenge raised: 
\begin{enumerate}
    \item {\bf Recognition of electron scattering as a necessary piece of a robust neutrino physics program} First, we summarize the motivation for new electron scattering measurements and theory as applied to neutrino physics (this white paper). The next step would be to commission impact metrics with experimental programs to refine needed precision and use of measurements.
    \item {\bf Expand, and use new funding opportunities} The Neutrino Theory Network will support proposals of theorists working on topics across NP and HEP. The formation of a topical group within DOE would avoid duplication of effort and review of related proposals.
\end{enumerate}

\section{Summary}
\label{sec:conclusions}


Neutrino physics has entered a precision era and
exciting neutrino experimental programs at low and high energies
can lead to discoveries.
The importance of constraining systematics resulted from neutrino-nucleus interaction physics in key neutrino measurements, in particular at accelerator-based experiments, cannot be overstated.  
The kinematically controlled beams and precise measurements of electron scattering data
makes it a key
 benchmark to assess and validate different nuclear models intended to be used in neutrino experiments. 

As part of the Snowmass 2021 process, we organized two dedicated Electron Scattering workshop: i) ``NF06 Electron Scattering''~\cite{NF06Workshop1} focused on long-baseline neutrino experiments, and ii) ``NF06 Low Energy Neutrino and Electron Scattering''~\cite{NF06Workshop2} focused on the low-energy (CEvNS and supernova) neutrinos that formed the basis of this contribution. In this White Paper, we highlight connections between electron- and neutrino-nucleus scattering physics at energies ranging from 10s of MeV to a few GeV, review the status of ongoing and planned electron scattering experiments, identified connections and gaps in the context of their input to the accelerator neutrino program. We also highlight the systemic challenges with respect to the divide between the nuclear and high-energy physics communities and funding that presents additional hurdle in mobilizing these connections to the benefit of neutrino programs.

While previous and existing electron scattering experiments provide important information, dedicated measurements and theoretical developments to support neutrino-nucleus interactions physics is needed. The NP-HEP cross community collective efforts will play a vital role in this endevour. The goal of these collective efforts will be to validate and solidify our understanding of the neutrino-nucleus interactions, enabling precision physics goals of neutrino experimental programs.



\section*{Acknowledgements}
\label{sec:acknowledgements}

We thank the speakers and participants of the ``NF06 Electron Scattering'' and ``NF06 Low Energy Neutrino and Electron Scattering'' Workshops~\cite{NF06Workshop1, NF06Workshop2} who have contributed both directly and indirectly to this White Paper.

The authors acknowledge the support of the U.S. Department of Energy (DOE) Office of Science, U.S. National Science Foundation (NSF), the Deutsche Forschungsgemeinschaft (DFG) through the Cluster of Excellence ``Precision Physics, Fundamental Interactions, and Structure of Matter" (PRISMA$^+$ EXC 2118/1) funded by the DFG within the German Excellence Strategy, the Research Foundation Flanders (FWO-Flanders), and the Spanish Ministerio de Economía y Competitividad (SEIDI-MINECO) under Grant No. PID2019-107564GB-I00, IFAE is partially funded by the CERCA program of the Generalitat de Catalunya.


\end{document}